\documentclass[prl,twocolumn,superscriptaddress,showpacs,amsmath,amstex,amssymb,citeautoscript]{revtex4-1}
\pdfoutput=1 

\usepackage{dcolumn}
\usepackage{bm}
\usepackage[T1]{fontenc}
\usepackage{braket}
\usepackage{graphicx}
\usepackage{xcolor}
\usepackage{lipsum}
\usepackage[normalem]{ulem}
\usepackage{amsmath}
\usepackage{amsfonts}
\usepackage{amssymb} 
\usepackage{color}
\usepackage[percent]{overpic}
\usepackage{soul} 
\usepackage{amssymb}
\usepackage{wasysym}
\usepackage{dsfont}
\usepackage{float}
\usepackage[mathscr]{euscript}
\usepackage{ifthen}
\usepackage{csquotes}

\begin{document}

\title{Spin transport in a tunable Heisenberg model realized with ultracold atoms}

\author{Niklas Jepsen$^{1,3}$, Jesse Amato-Grill$^{1,3}$, Ivana Dimitrova$^{1,3}$, Wen Wei Ho$^{2,3}$, Eugene Demler$^{2,3}$ \& Wolfgang Ketterle$^{1,3}$}
\noaffiliation

\affiliation{Department of Physics and Research Laboratory of Electronics, Massachusetts Institute of Technology, Cambridge, Massachusetts 02139, USA}
\affiliation{Department of Physics, Harvard University, Cambridge, Massachusetts 02138, USA}
\affiliation{MIT-Harvard Center for Ultracold Atoms\vspace{-10pt}}
	 
\maketitle

\noindent 
\textbf{Simple models of interacting spins play an important role in physics. They capture the properties of many magnetic materials, but also extend to other systems, such as bosons and fermions in a lattice, systems with gauge fields, high-$\mathbf{T_c}$ superconductors, and systems with exotic particles such as anyons and Majorana fermions. In order to study and compare these models, a versatile platform is needed. Realizing such a system has been a long-standing goal in the field of ultracold atoms. So far, spin transport has only been studied in the isotropic Heisenberg model \cite{munich_singleSpin, munich_boundMagnons, munich_spinHelix, zwierlein_spinTransport}. Here we implement the Heisenberg XXZ model with adjustable anisotropy and use this system to study spin transport far from equilibrium after quantum quenches from imprinted spin helix patterns. In the non-interacting XX model, we find ballistic behavior of spin dynamics, while in the isotropic XXX model, we find diffusive behavior. For positive anisotropies, the dynamics  ranges from  anomalous super-diffusion to sub-diffusion depending on anisotropy, whereas for negative anisotropies, we observe a crossover in the time domain from ballistic to diffusive transport. This behavior contrasts with expectations for the linear response regime and raises new questions in understanding quantum many-body dynamics far away from equilibrium.}
    
Quantum dynamics is an active frontier of many-body physics, as it underlies a myriad of physical phenomena such as transport, thermalization, and novel nonequilibrium states of matter. However, even the linear response (near equilibrium) behavior of many-body systems can be very complex. For example, spin transport in integrable one-dimensional Heisenberg XXZ quantum spin chains, despite being a topic that is decades old, is still under active  investigation \cite{3_Vasseur_2016, 4_Ljubotina:2017aa, 5_PhysRevLett.122.127202, 6_PhysRevLett.121.230602}. Dynamics in highly out-of-equilibrium scenarios, such as arising from continual drives or quantum quenches \cite{Choi_2017, Zhang_2017_DTC, Cavalleri_2018, Basov_2017, Kampfrath_2013, Berges_2004,Gring_2012, Pruefer_2018, Eigen_2018, Bernien_2017, Zhang_2017}, is even less well understood. It is hence highly desirable to have a quantum simulator that can realize well-isolated, programmable and controllable quantum many-body systems in different scenarios. By now, a number of such platforms exist \cite{Barends2015, Bernien_2017, Zhang_2017, Gross995, ferlaino20, bakr18, yan13, browaeys20, moses17, superconductingcubits20, SchleierSmith_Heisenberg, superexchange, zwierlein_fermiMottInsulator,   greiner_fermiSpinCorrelations, greiner_fermiAntiferromagnet, chiu19, brown15}, with varying capabilities,  allowing for the systematic  study of a variety of interesting quantum many-body dynamical phenomena.

Ultracold atoms in optical lattices offer an especially promising platform to realize isolated, well-controlled and tunable Heisenberg spin models \cite{zwierlein_fermiMottInsulator,   greiner_fermiSpinCorrelations, greiner_fermiAntiferromagnet, chiu19, brown15}. Indeed, in deep lattices where atoms become localized on individual sites hence forming a Mott insulator \cite{boseHubbardModel, jakschboseHubbardModel}, the dynamics of the remaining degrees of freedom is described by effective spin-spin interactions, thereby realizing nearest-neighbor Heisenberg XXZ spin models. For bosons, the most commonly used atom, $^{87}$Rb, has almost equal singlet and triplet scattering lengths, implying effectively isotropic spin physics \cite{superexchange, munich_singleSpin, munich_boundMagnons, munich_spinHelix}. For fermions, the Pauli exclusion principle enforces isotropic anti-ferromagnetism \cite{zwierlein_spinTransport, zwierlein_fermiMottInsulator, greiner_fermiSpinCorrelations, greiner_fermiAntiferromagnet}. Although many theoretical proposals have suggested ways to obtain richer spin models \cite{counterflowSF, duanDemlerLukin, GarciaRipollCirac, AltmanHofstetterDemlerLukin}, it is only now, almost twenty years later, that we report here the realization of a spin-1/2 Heisenberg model with adjustable anisotropy in the spin-spin interactions. This wide tunability is realized using $^7$Li atoms, whose Feshbach resonances we have characterized in our previous work \cite{interactionSpectroscopy}. Additionally, because the spin-spin couplings are mediated by second-order tunneling (superexchange) \cite{superexchange}, using lithium with its light mass leads to fast spin dynamics, decreasing the relative importance of heating and loss processes compared to using  heavier atoms.

For many-body quantum simulation experiments, one ideally starts by implementing a simple benchmark system and then adds interactions, realizing more complex and less well-understood systems. In this work, we first implement the XX model in 1D, which is exactly solvable and can be mapped to a system of non-interacting fermions by the Jordan-Wigner transformation \cite{JordanWigner}. We then tune the anisotropy to arbitrary values, which in the fermionic language corresponds to adding nearest-neighbor interactions, which can be attractive or repulsive.

\begin{figure}[t]
	\includegraphics[width=\linewidth]{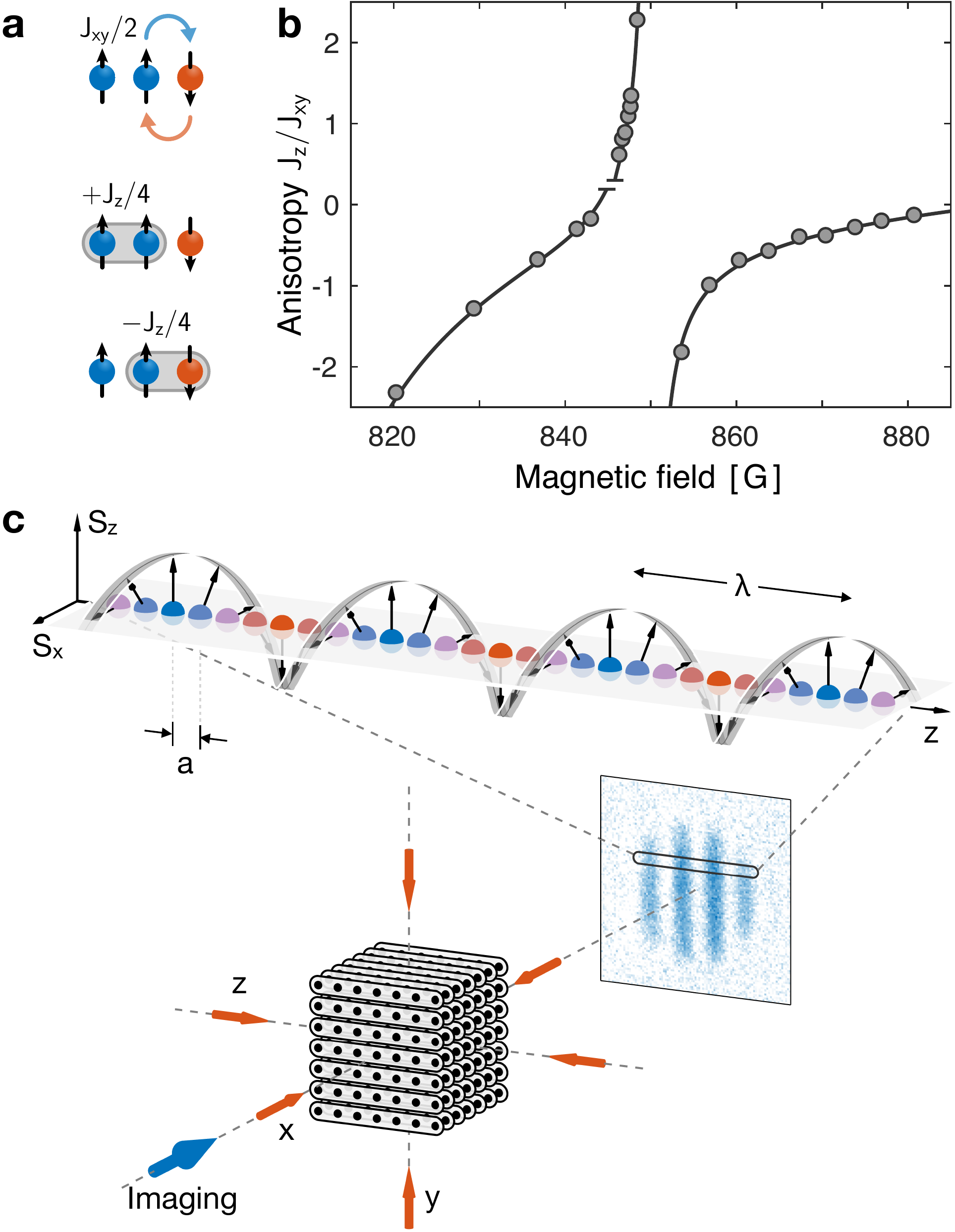}
	\caption{
		\textbf{Tunability of the XXZ model with $^7$Li and implemenation of a spin helix.}  \textbf{a}, The Hamiltonian \eqref{Heisenberg_eq} is characterized by two energy scales: Transverse spin coupling $ J_{xy} $ (spin-exchange) and longitudinal spin coupling $ J_z $ (spin-spin interactions). 
		\textbf{b}, Anisotropy $\Delta\,{=}\,J_z/J_{xy}$ as a function of applied magnetic field. The solid line is a fit to experimental data points which are calculated from measured values $ U_{\uparrow\uparrow} $, $ U_{\uparrow\downarrow} $, $ U_{\downarrow\downarrow}$ (see Methods). 
		\textbf{c}, Spin helix realized from two hyperfine states (spin $|{\uparrow}\rangle$ and $|{\downarrow}\rangle$). The spin ${\bf S}$ winds within the $S_z S_x$-plane as a function of position $z$ in the spin chain. Deep optical lattices along $x$ and $y$ create an array of independent spin chains. The $z$-lattice is shallower and controls spin transport along each chain.}
	\label{fig:setup}
	\vspace{-18pt}
\end{figure}

To implement the spin model, we use a system of two-component bosons in an optical lattice, which is well-described by the Bose-Hubbard model. These two states can be identified with a spin-$1/2$ degree of freedom $ \ket{\uparrow} $ and $ \ket{\downarrow} $, and in the Mott insulating regime at unity filling the effective Hamiltonian is given by the spin-1/2 Heisenberg XXZ model \cite{counterflowSF, duanDemlerLukin, GarciaRipollCirac, AltmanHofstetterDemlerLukin} 
\begin{equation}
	H = \sum_{\langle i j \rangle} \left[ J_{xy} (S_i^x S_j^x + S_i^y S_j^y) + J_z S_i^z S_j^z \right]
    \label{Heisenberg_eq}
\end{equation}
where nearest-neighbor $\langle ij\rangle$ couplings are mediated by superexchange. Here, in leading order, $J_{xy}\,{=}\,{-}\,4t^2/U_{\uparrow\downarrow}$ and $J_z\,{=}\,4t^2/U_{\uparrow\downarrow}\,{-}\,(4t^2/U_{\uparrow\uparrow}\,{+}\,4t^2/U_{\downarrow\downarrow})$,
where $t$ is the tunneling matrix element between neighbouring sites, and $U_{\uparrow\uparrow}$, $U_{\uparrow\downarrow}$, $U_{\downarrow\downarrow}$ are the on-site interaction energies. The transverse coupling $ J_{xy} $ induces spin-exchange between neighboring sites and is the origin of spin transport. The longitudinal coupling $ J_z $ corresponds to a nearest-neighbor spin-spin interaction (Fig.~\ref{fig:setup}a).

The magnitude of superexchange can be varied over two orders of magnitude by changing the lattice depth, which scales the entire Hamiltonian. Due to axial symmetry, the total magnetization $\sum_i S^z_i$ is conserved. We control the anisotropy $\Delta\,{:=}\,J_z/J_{xy}$ via an applied magnetic field which tunes the interactions through Feshbach resonances in the lowest two hyperfine states (Fig.~\ref{fig:setup}b). In the regime studied here, the transverse coupling is positive ($J_{xy}\,{>}\,0$). The ability to tune the anisotropy over a wide range of positive and negative values allows us to explore dynamics beyond previous experiments in which $\Delta\,{\approx}\,1$ \cite{munich_singleSpin, munich_boundMagnons, munich_spinHelix,zwierlein_spinTransport, zwierlein_fermiMottInsulator, greiner_fermiSpinCorrelations, greiner_fermiAntiferromagnet}.

In this experiment, 1D chains are implemented by two perpendicular optical lattice beams whose depths $V_x$,$V_y\,{=}\,35\,E_R$ are sufficient to prevent tunneling in the $x$- and $y$-direction on experimental timescales, and a third mutually orthogonal lattice beam whose depth $V_z$ controls the superexchange rate in the chains along the $z$-direction (Fig.~\ref{fig:setup}c). Here $E_R\,{=}\,{h^2/(8ma^2)}$ denotes the recoil energy, where $a$ is the lattice spacing and $m$ the atomic mass. After preparing an identical spin helix with wavelength $\lambda$ in each chain (see Methods), time evolution is initiated by rapidly lowering $V_z$. The coherent dynamics following this quench is therefore governed in each chain by the 1D XXZ model (Eq.~\eqref{Heisenberg_eq}) with an anisotropy $\Delta$ selected by an appropriate applied magnetic field. After an evolution time $t$ of up to $500\,\hbar/J_{xy}$ (well below the heating lifetime ${\sim1}\,s $ of the Mott insulator), the dynamics is frozen by rapidly increasing $V_z$ and the atoms are imaged in the $\ket{\uparrow}$ state via state-selective polarization-rotation imaging with an optical resolution of about 6 lattice sites (see Methods).

Integrating the images along the direction perpendicular to the chains yields a 1D spatial profile of the population in the $\ket{\uparrow}$ state, averaged over all spin chains (see Extended Data Fig.~\ref{ED-fig:imageFit}). This is equivalent to a measurement of the expectation value of the local magnetization ${\braket{S^z_i}}\,{=}\,{(n_{i,\uparrow}\,{-}\,n_{i,\downarrow})/2}\,{=}\,{n_{i,\uparrow}\,{-}\,1/2}$. As in Fig.~\ref{fig:setup}c, the initial spin helix exhibits a maximum-contrast sinusoidal spatial modulation of $\braket{S^z_i}$, which results in a characteristic stripe pattern after imaging. We determine the contrast $\mathcal{C}$ after an evolution time $t$ by a fit $f(z)\,{=}\,{g(z)\,{\cdot}\,[{1}\,{+}\,{\mathcal{C} \cos(Qz+\theta)}]/2}$, where $Q\,{=}\,2\pi/\lambda$ is the wavevector, $g(z)$ is a Gaussian envelope function which accounts for the spatial distribution of all atoms $n\,{=}\,n_\uparrow\,{+}\,n_\downarrow$, and $\theta$ is a random phase which varies from shot to shot due to small magnetic bias field drifts. During the evolution time $t$ this contrast $\mathcal{C}(t)$ decays, and we determine the dependence of $c(t)\,{=}\,\mathcal{C}(t)/\mathcal{C}(0)$ on lattice depth, wavelength $\lambda$, and anisotropy $\Delta$.

\begin{figure*}[t]
    \includegraphics[width=\linewidth,keepaspectratio]{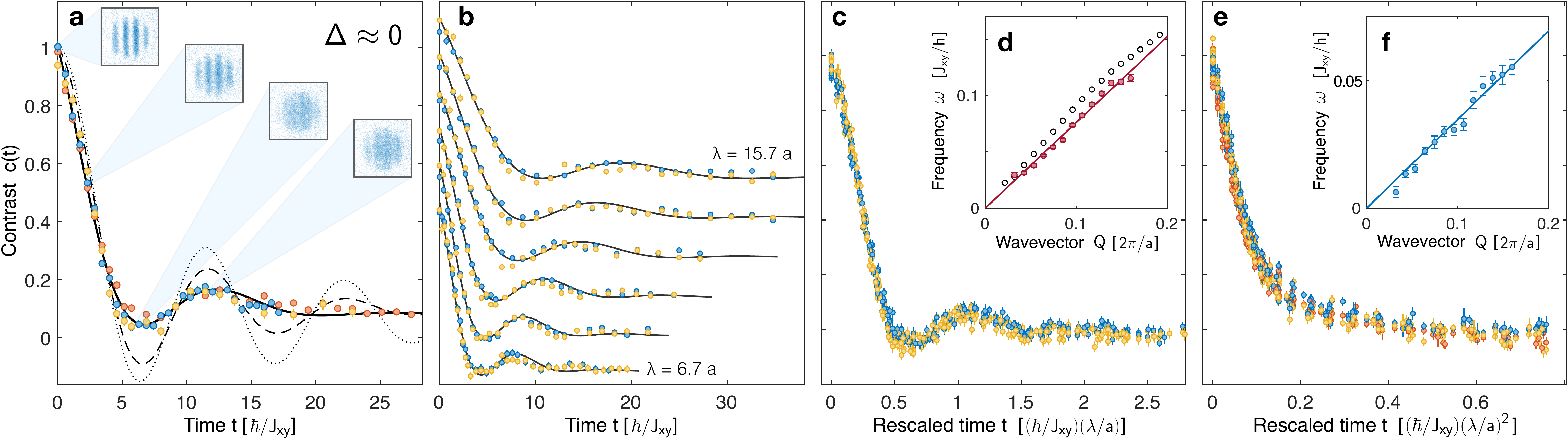}
    \caption{
        \textbf{Ballistic and diffusive spin transport}. \textbf{a}-\textbf{d}, XX model, ballistic behavior ($\Delta\,{\approx}\,0$, non-interacting fermions). \textbf{e}-\textbf{f}, XXX model, diffusive behavior ($\Delta\,{\approx}\,1$, strongly-interacting fermions). \textbf{a}, Spin helix contrast $c(t)$ for $\lambda\,{=}\,10.4\,a$ measured at three different lattice depths. The curves collapse when times are rescaled in units of $\hbar/J_{xy}\,{=}\,0.75\,\text{ms}$, $2.01\,\text{ms}$, $5.08\,\text{ms}$ for lattice depths of $9\,E_R$ (red), $11\,E_R$ (blue), $13\,E_R$ (yellow). The fit (black line) shows a decay with time-constant $\tau\,{=}\,5.5(2)\,\hbar/J_{xy}$ and a damped oscillation with period $T\,{=}\,2\pi/\omega\,{=}\,13.7(2)\,\hbar/J_{xy}$. Numerical simulations are also shown for the XX model (dotted line) and bosonic $t$-$J$-model with $5\,\%$ holes (dashed line). \textbf{b}, Decay curves for different wavelengths $\lambda\,{=}\,15.7\,a$, $13.4\,a$, $11.7\,a$, $9.4\,a$, $7.8\,a$, $6.7\,a$ (offset for clarity) can be collapsed into a single curve, \textbf{c}, if time units are rescaled by $\lambda$ (indicating ballistic transport) and offsets $c_0$ are removed. \textbf{d}, The oscillation frequencies (filled symbols) follow a linear dispersion relation $\omega(Q)$ with a velocity $v\,{=}\,0.76(1)\,v_F$ and are in agreement with numerical simulations (open symbols) yielding $v\,{=}\,0.85(1)\,v_F$. Theoretical frequencies are obtained as the inverse of the first revival time.  Due to damping, this may overestimate frequencies by $10\,\%$. \textbf{e}, For $\Delta\,{\approx}\,1$, oscillations are strongly suppressed and time units have to be rescaled by $\lambda^2$ (indicating diffusive transport). However this collapse is not perfect, because, \textbf{f}, the small oscillations are still slightly visible and follow a linear dispersion relation $\omega(Q)$. The velocity $v\,{=}\,0.35(1)\,v_F$ is more than a factor of 2 smaller than in the non-interacting case $\Delta\,{\approx}\,0$. (Also see Extended Data Fig.~\ref{ED-fig:dispersion}) }
    \label{fig:XX}
	\vspace{-0pt}
\end{figure*}

For all data, we measure the spin dynamics at two or three different lattice depths $V_z$ and verify that the decay curves $c(t)$ collapse when time is rescaled by the corresponding spin-exchange time $\hbar/J_{xy}$ (see e.g.~Fig.~\ref{fig:XX}a). This demonstrates that what we observe is transport by superexchange and not some other process, such as transport of defects (for which timescales would scale linearly with the tunneling matrix element). Throughout the rest of the paper, time is normalized by the spin-exchange time $\hbar/J_{xy}$, length by the lattice spacing $a$ and velocities are expressed in units of the Fermi velocity $v_F\,{=}\,a/(\hbar/J_{xy})$. These units are obtained from the experimentally determined lattice depth using an extended Hubbard model and have an estimated systematic calibration error of about $\pm10\,\%$, in addition to quoted statistical errors. The accuracy of the experimental calibration of the anisotropy $\Delta$ is estimated to be ${\pm 0.1}$ (see Methods).

{\it XX model.} We first study the case $\Delta\,{=}\,0$, which can be mapped by the Jordan-Wigner transformation \cite{JordanWigner} to non-interacting spinless fermions undergoing nearest-neighbor hopping on a lattice. In this mapping, spin $\ket{\uparrow}$ corresponds to a site occupied by a fermion $\ket{1}$, and $\ket{\downarrow}$ to an empty site $\ket{0}$. The band structure is $E(q)\,{=}\,J_{xy}\cos(qa)$ where $q$ is the lattice momentum. For equal number of $ \ket{\uparrow}$ and $ \ket{\downarrow}$, the ground state corresponds to a half-filled band. Small excitations around this Fermi sea are spin waves with a linear dispersion relation $\omega(q)\,{=}\,v_F q$, where $v_F$ is the Fermi velocity.

Fig.~\ref{fig:XX}a shows the decay of the contrast $c(t)$ for $\Delta\,{\approx}\,0$. (The data was taken for $\Delta\,{=}\,0.02 $ according to lowest-order Hubbard parameters. However, higher-order corrections (see Methods) result in an actual value of $\Delta\,{=}\,{-0.12}$. Since the results in this regime are only weakly dependent on $\Delta$, we refer to those measurements as $\Delta\,{\approx}\,0$.) In addition to an overall decay, a local maximum corresponding to a partial revival of the initial spin modulation appears after about 12 spin-exchange times. We find the decay curves can be well described by the sum of a decaying part with time constant $\tau$ and a (damped) oscillating part with frequency $\omega$, resulting in a fitting function $c(t)\,{=}\,{\left( a_0\,{+}\,b_0\cos\omega t\right) e^{-t/\tau}}\,{+}\,c_0$. Here $a_0$, $b_0$, $c_0$, $\omega$, $\tau$ are fitting parameters (for discussion on the constant offset $c_0$ see Methods and Extended Data Fig.~\ref{ED-fig:holes}). Numerical simulations, also shown in Fig.~\ref{fig:XX}a, agree qualitatively with the experimentally observed dynamics. By varying the wavelength $\lambda\,{=}\,2\pi/Q $ of the helix (Fig.~\ref{fig:XX}b) we obtain a dispersion relation $\omega(Q)$ for the oscillations (Fig.~\ref{fig:XX}d). A linear fit $\omega(Q)\,{=}\,vQ$, yields a characteristic velocity $v\,{=}\,0.76(1)\,v_F$, a value remarkably close to the Fermi velocity at half filling, even though our initial condition (spin helix) is a state far from equilibrium and hence not necessarily governed by (near-equilibrium) spin waves.

\begin{figure}[t]
    \includegraphics[width=1\linewidth,keepaspectratio]{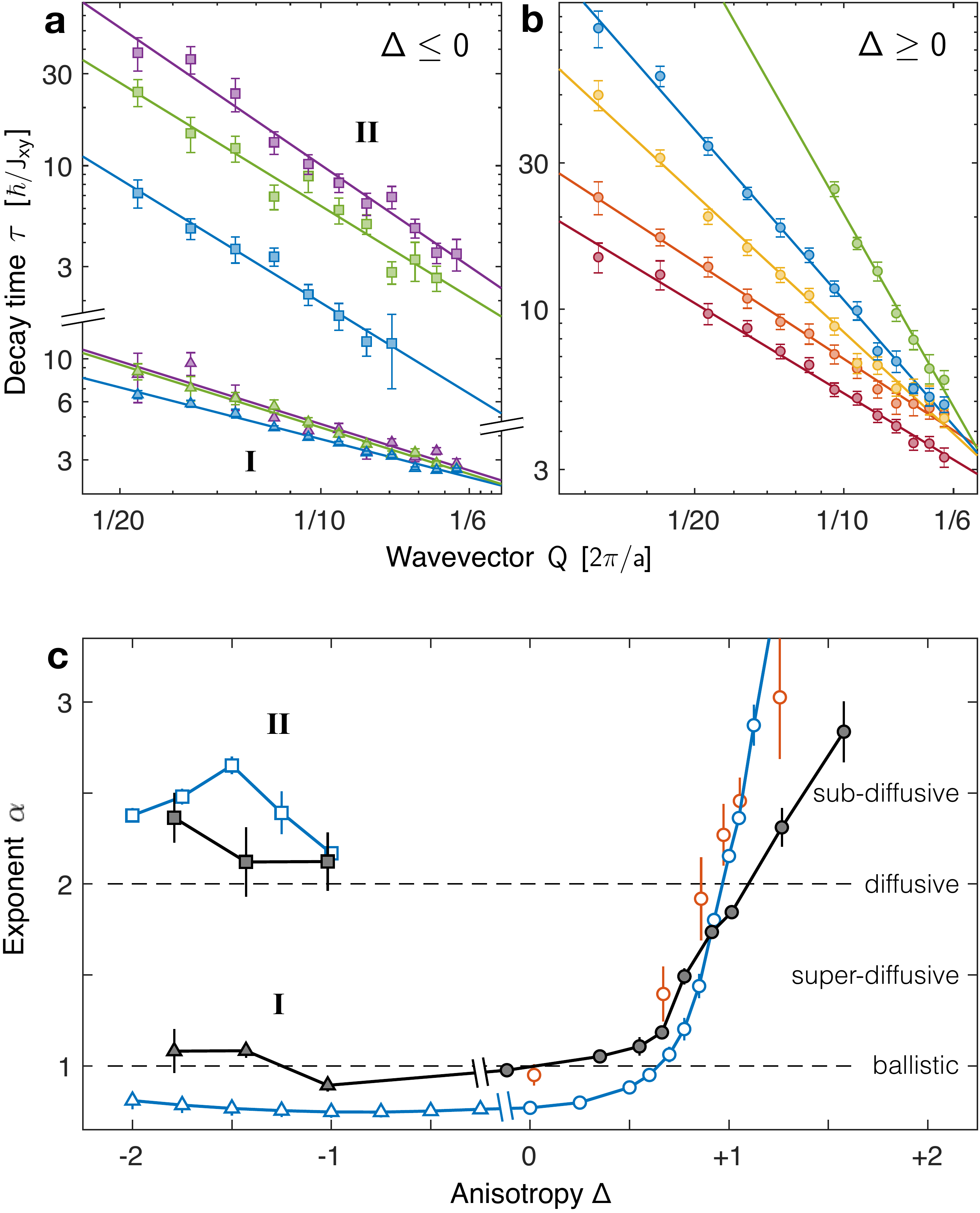}
    \caption{
        \textbf{Power law scalings} of decay time constants $\tau$ for different anisotropies $\Delta$ ranging from (\textbf{a}) negative to (\textbf{b}) positive. Experimental results are shown in \textbf{a} for $\Delta\,{=}\,{-1.02}$ (blue), ${-1.43}$ (green), ${-1.79}$ (purple) and in \textbf{b} for $\Delta\,{=}\,{-0.12}$ (red), ${0.35}$ (orange), ${0.78}$ (yellow), ${1.01}$ (blue), ${1.58}$ (green). Lines are power law fits. \textbf{c}, Fitted power law exponents from experiments (filled symbols) and theory (open symbols), obtained from simulations of the defect-free XXZ model (blue) and $t$-$J$ model with $5\,\%$ hole fraction (red). \textbf{a}-\textbf{c}, For $\Delta\,{\geq}\,0$ (circles) we observe anomalous diffusion: The exponent increases smoothly from ballistic (red) to super-diffusive (yellow) to diffusive (blue) to sub-diffusive (green). For $\Delta\,{<}\,0$ we observe behavior reminiscent of a classical gas: Transport is ballistic at (I) short times (triangles) and diffusive at (II) longer times (squares). (For diffusion coefficients and theory see Extended Data Fig.~\ref{ED-fig:powerlaw_theory}) }
	\label{fig:powerlaw}
	\vspace{-15pt}
\end{figure}

The decay time constant $\tau$ also shows a linear scaling with inverse wavevector. Specifically, a power law fit $\tau\,{\propto}\,Q^{-\alpha}$ yields an exponent of $\alpha\,{=}\,0.98(3)$ consistent with 1, indicating ballistic transport (Fig.~\ref{fig:powerlaw}b, red). Indeed, if we plot $c(t)$ in time units rescaled by the wavelength $\lambda$, then all curves for different helix wavelengths collapse into a single curve (Fig.~\ref{fig:XX}c) with a decay time $\tau\,{=}\,0.53(1)\,\lambda/v_F$ and an oscillation period of $T\,{=}\,2\pi/\omega\,{=}\,1.31(2)\,\lambda/v_F$. This collapse shows that all aspects of the observed spin dynamics in the XX model are, for all evolution times $t$, ballistic and governed by one characteristic velocity.

{\it XXX model.} For finite $\Delta$, the Jordan-Wigner transformation results in fermions with nearest-neighbor interactions. The isotropic case $\Delta\,{=}\,1$ corresponds to strong interactions, which should generically turn fast ballistic transport into slow diffusive transport. Indeed, the decay slows down for increasing wavelength $\lambda$ much more dramatically than in the $\Delta\,{\approx}\,0$ case (also illustrated in Extended Data Fig.~\ref{ED-fig:dispersion}b). A power law fit of the decay constants $\tau$ versus $Q$ yields an exponent of $\alpha\,{=}\,1.85(4)$, which is close to 2, indicative of a diffusive process (Fig.~\ref{fig:powerlaw}b, blue). If we rescale time units by $\lambda^2$, then all contrast curves $c(t)$ collapse very well into a single curve (Fig.~\ref{fig:XX}e).  However this collapse is not perfect, because we can observe a (small) oscillating part which still obeys a linear dispersion relation $\omega(Q)\,{=}\,vQ$ (Fig.~\ref{fig:XX}f).
 
Using the relation $1/\tau\,{=}\,DQ^2$, a diffusion constant can be determined as $D\,{=}\,0.242(7)\,{a^2/(\hbar/J_{xy})}$. Interpreting $D\,{=}\,\frac{1}{2} \delta x^2 / \delta t$ as a random walk of step size $\delta x$ (a \enquote{mean free path}) and time $\delta t$ between steps, and using $v\,{=}\,\delta x/\delta t\,{=}\,0.35(1)\,v_F$ (obtained from the dispersion relation in Fig.~\ref{fig:XX}f), we find $\delta x\,{=}\,1.38(6)\,a$. A mean free path on the order of the lattice constant is analogous to the Ioffe-Regel limit for resistivity where simple quasi-particle pictures break down \cite{Ioffe,Mott}, implying that the isotropic Heisenberg model is strongly interacting.  

Our observation of diffusive behavior and our measured value for the diffusion coefficient are consistent with previous work \cite{munich_spinHelix} on the 1D isotropic Heisenberg model, which was done for a ferromagnetic system ($J_{xy}\,{<}\,0$). Since our system is antiferromagnetic ($J_{xy}\,{>}\,0$), this indicates that the overall sign of the Hamiltonian is irrelevant, as expected from theoretical arguments involving time-reversal symmetry (see Methods). We note, however, that the small (ballistic) oscillatory component has not been previously observed.

By tuning the interactions over a large range of $\Delta$, we study how the transport behavior changes. For an interacting gas of classical particles or quasiparticles, one would expect ballistic behavior on timescales shorter than the collision time and diffusion for longer times.  That is what we find for $\Delta\,{<}\,0$, whereas for $\Delta\,{\geq}\,0$ we observe qualitatively very different behaviour (also shown in Extended Data Fig.~\ref{ED-fig:decay_anisotropies}).

{\it Positive anisotropies} ($\Delta\,{\geq}\,0$). All measured temporal decay curves $c(t)$ for positive anisotropies are well described by the fitting function previously used, and the observed scaling of time-constants are shown in Fig.~\ref{fig:powerlaw}. Over the full range of investigated wavevectors $Q$, we find the  decay time constants $\tau$ obey power laws $\tau\,{\propto}\,Q^{-\alpha}$ (solid lines) in the following way: as the anisotropy is increased from $\Delta\,{=}\,{-0.12}$ to $\Delta\,{=}\,0.55$, the exponent stays close to $\alpha\,{=}\,1$ (\enquote{ballistic regime}), but the characteristic velocity of oscillations decreases by a factor of about 1.5 to $v\,{=}\,0.47(1)\,v_F$. Between $\Delta\,{\approx}\,0.55$ and $ 1 $ the exponent increases smoothly from $\alpha\,{=}\,1 $ to $\alpha\,{\approx}\,2$ (\enquote{super-diffusive regime}). For example, we measure $\alpha\,{=}\,1.49(4)$ at $\Delta\,{=}\,0.78$ (Fig.~\ref{fig:powerlaw}b, yellow). For $\Delta\,{>}\,1$ transport slows down even more, and the exponent also continues to increase smoothly to values $\alpha\,{>}\,2$ (\enquote{sub-diffusive regime}). For example, at $\Delta\,{=}\,1.58$, $\alpha\,{=}\,2.84(17)$ (Fig~\ref{fig:powerlaw}b, green). For each positive anisotropy $\Delta$, the measured decay curves for different wavelengths $\lambda$ collapse into a single curve, if time units are rescaled by $\lambda^\alpha$ (Extended Data Fig.~\ref{ED-fig:universal}).

Power law exponents between 1 and 2 (super-diffusion) are often associated with L\'{e}vy flights or fractional Brownian motion where step sizes are correlated \cite{DUBKOV_2008, Kolmogorov, Yaglom}. Power law exponents larger than 2 (sub-diffusion) typically arise for transport through a disordered medium \cite{Agarwal_2015, Vosk_2015} and have also been recently observed in a tilted Fermi-Hubbard system \cite{bakr20}. However, there is no disorder in the XXZ Heisenberg Hamiltonian we study. We note that even in the sub-diffusive regime the fit still finds a small oscillatory component whose frequency obeys a linear dispersion relation $\omega(Q)\,{=}\,vQ$ (Extended Data Fig.~\ref{ED-fig:dispersion}). For example, at $\Delta\,{=}\,1.27$ we measure an exponent of $\alpha\,{=}\,2.31(11)$ (sub-diffusive) and a characteristic velocity of $v\,{=}\,0.29(1)\,v_F$. This ballistic component may be related to our initial condition of a spin helix which in the mapping to lattice fermions is a $100\,\%$ density modulation, which reduces scattering at early times.

\begin{figure*}[t!]
    \includegraphics[width=0.75\linewidth,keepaspectratio]{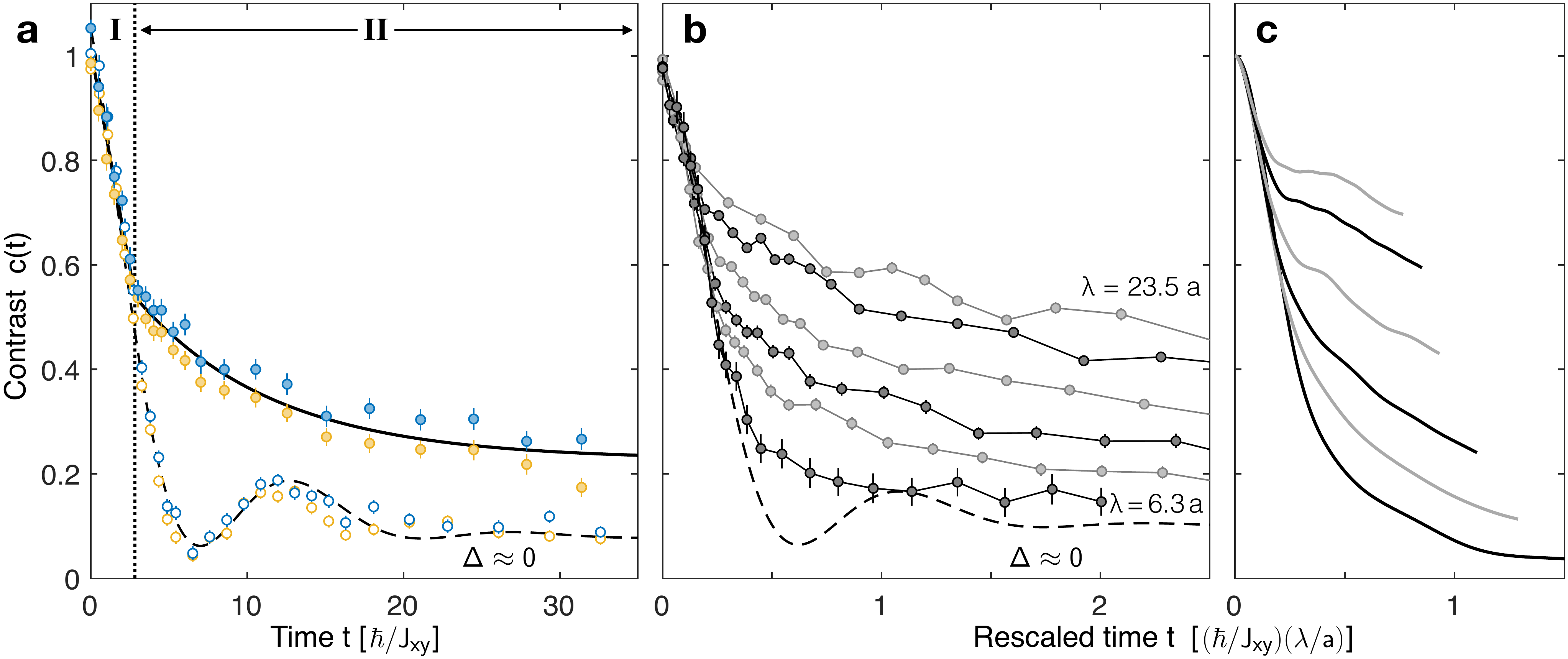}
    \caption{
        \textbf{Temporal crossover from ballistic to diffusive transport} for negative anisotropies $\Delta\,{<}\,0$. \textbf{a}, Spin helix contrast $c(t)$ for $\lambda\,{=}\,10.4\,a$ and $\Delta\,{=}\,{-1.43}$ (filled circles) measured at two different lattice depths $11\,E_R$ (blue) and $13\,E_R$ (yellow). A piecewise fit (solid line) is linear at short times (I) and exponential at long times (II) with a sharp crossover at $t\,{=}\,t_0$ (vertical dotted line). For $t\,{<}\,t_0$ the decay coincides well with the non-interacting case $\Delta\,{\approx}\,0 $ (open circles and dashed line). \textbf{b}, Decay curves for different wavelengths $\lambda\,{=}\,23.5\,a$, $18.8\,a$, $13.4\,a$, $10.4\,a$, $8.5\,a$, $6.3\,a$ (average of $11\,E_R$ and $13\,E_R$) collapse into a single curve at early times, if time units are rescaled by $\lambda$ (as for ballistic behaviour) and follow the non-interacting case (dashed line). At later times, the decay is diffusive with different scaling (see Fig.~\ref{fig:powerlaw}a). \textbf{c}, Numerical simulations for $\Delta\,{=}\,{-1.5}$ and the same wavelengths $\lambda$ as in panel \textbf{b} show similar behavior. Our simulations could not be extended to longer times due to exponential increase in computation time.}
    \label{fig:negative}
\end{figure*}

{\it Negative anisotropies} ($\Delta\,{<}\,0$). Here the behavior is qualitatively very different compared to the positive cases for similar $|\Delta|$. We find a crossover in the time domain from ballistic to diffusive behavior. For example, at $\Delta\,{=}\,{-1.43}$ the initial decay of the contrast $c(t)$ is fast and, in fact, coincides well with the non-interacting (ballistic) case $\Delta\,{\approx}\,0$  (Fig.~\ref{fig:negative}a), in stark contrast to the positive case $\Delta\,{=}\,{+1.58}$ with similar magnitude (Extended Data Fig.~\ref{ED-fig:decay_anisotropies}). At $t\,{=}\,t_0\,{\approx}\,2.8\,\hbar/J_{xy}$ (dotted line) the decay suddenly slows down. We therefore parameterize the decay curve $c(t)$ by a piecewise fit with two timescales: (I) a linear function $(1\,{-}\,t/\tau_\text{I})$ at short times and (II) an exponential $e^{-t/\tau_\text{II}}$ at longer times, with respective time constants $\tau_\text{I}$, $\tau_\text{II}$. When the wavevector $Q$ is varied, both $\tau_\text{I}$ and $\tau_\text{II}$ follow a power law (Fig.~\ref{fig:powerlaw}a), but with different exponents: $\alpha_\text{I}\,{=}\,1.08(4)$ (ballistic) and $\alpha_\text{II}\,{=}\,2.12(19)$ (diffusive)  respectively (Fig.~\ref{fig:powerlaw}c). In both the experimental (Fig.~\ref{fig:negative}b) and the numerical results (Fig.~\ref{fig:negative}c), all decay curves collapse for short times before \enquote{peeling off} at later times, if time units are rescaled by $\lambda$.

Similar behavior is observed for other negative anisotropies (Fig.~\ref{fig:powerlaw}a; also Extended Data Fig.~\ref{ED-fig:negative_anisotropies}), with the initial ballistic temporal decay (regime I) almost independent of $\Delta$. However, for larger $|\Delta|$, we find that $t_0$ (the range of regime I) is smaller, and the diffusion timescales $\tau_\text{II}$ in regime II are longer.  The diffusion coefficient decreases from $D\,{=}\,1.27(6)$ to $0.25(2)\,a^2/(\hbar/J_{xy})$ when the interactions are increased from $\Delta\,{=}\,{-1.02}$ to ${-1.79}$ (Extended Data Fig.~\ref{ED-fig:powerlaw_theory}). Fig.~\ref{fig:powerlaw}c summarizes the different transport behaviors we have discovered for the anisotropic Heisenberg model and it represents the main result of this paper.

{\it Theoretical simulations.} In order to validate our platform as a quantum simulator, we have carried out numerical simulations of quench dynamics starting from a spin helix, using a combination of exact diagonalization and tensor network methods (see Methods). We simulate the dynamics of the system without holes (XXZ Hamiltonian), as well as with a small probability of holes (bosonic $t$-$J$ model), and compare the simulated contrast to that measured in the experiments.  

The time scales of decay in simulations and experiments generally agree fairly well.  A qualitative difference in the decay curves is illustrated in Fig.~\ref{fig:XX}a: in the simulations, there is always an initial quadratic decay component (as expected from time-reversal symmetry of the system, see Methods), and revivals are generally more pronounced. The initial quadratic decay happens in the pure-spin simulations on the timescale of $\hbar/J_{xy}$, while an addition of $2.5\,\%$ to $5\,\%$ holes reduces this to the timescale $\hbar/t$, where $t$ is the hopping amplitude in the $t$-$J$ model (see also \cite{munich_spinHelix}), and reduces the amplitude of revivals. However, the presence of holes does not affect the overall decay times of the spin contrast: the simulations of both the XXZ and the $t$-$J$ model yield power law scalings of time constants with exponents which agree reasonably well with the experimental ones (Fig.~\ref{fig:powerlaw}c and Methods).

{\it Discussion.} Our work on spin transport illustrates the strength of a combined experimental and theoretical quantum simulation and reveals current limitations: even for small 1D systems, computational resources soon reach limits regarding chain length and time steps. In general, experimental decay curves were simpler to parametrize and showed better power law scaling than simulations. In some examples we could show that this is due to ensemble averaging in the experiment (chains of different lengths, different initial phases of the helix, and the presence of holes).  Simulating these effects was often computationally prohibitive. On the other hand, simulations provided valuable insight into the effects of holes (Figs.~\ref{fig:XX}a, \ref{fig:powerlaw}c) and the role of boundary conditions (Extended Data Figs.~\ref{ED-fig:initial_phase}, \ref{ED-fig:chain_lengths}), which could not be studied experimentally.

Our observations are consistent with some theoretical predictions for spin transport in the anisotropic Heisenberg model, but at the same time differ sharply from others. For example, studies of quantum quenches from pure states involving a single domain wall \cite{PhysRevE.59.4912, PhysRevE.71.036102, 10.21468/SciPostPhys.7.2.025} have suggested ballistic dynamics at $\Delta\,{=}\,0$ and diffusive dynamics (albeit with logarithmic corrections) at $\Delta\,{=}\,1$, similar to our findings. In contrast to our findings, theoretical studies of long-time, linear-response of spin transport at high temperatures (i.e.~mixed states) have indicated that the transition from ballistic to diffusive transport as a function of anisotropy is sharp: $\alpha\,{=}\,1$ for $\Delta\,{<}\,1$, $\alpha\,{=}\,3/2$ at $\Delta\,{=}\,1$, while $\alpha\,{=}\,2$ for $\Delta\,{>}\,1$ \cite{4_Ljubotina:2017aa, 5_PhysRevLett.122.127202}, which can be understood in a recently developed theoretical framework of generalized hydrodynamics involving local equilibriation of conserved quantities \cite{GHD1, GHD2}. We stress that the situation we have considered -- quenches from spin helix states far from equilibrium -- is different, and a direct comparison may not be possible. An accurate analysis of coherent dynamics starting from the initial helix state is a very challenging many-body problem since it cannot easily be represented in terms of the exact eigenstates of the model using the Bethe ansatz \cite{Caux_2011}. The rich phenomenology observed in our experiments and dramatic differences with the cases studied in the literature calls for a deeper understanding of this new dynamical regime, both theoretically and experimentally.

In conclusion, we have used ultracold atoms to implement Heisenberg spin models in a highly controlled and tunable way, utilizing Feshbach resonances to vary the anisotropy. We have studied far-from-equilibrium spin transport and explored how transport slows down from ballistic behavior (for a non-interacting system in fermionic language) to slow diffusion when interactions were introduced and found qualitatively different behavior for attractive and repulsive interactions.  Our combined experimental and theoretical studies have uncovered unexpected spin dynamics in regimes far from previously studied. This demonstrates the power of our platform as a quantum simulator to study general spin physics in new regimes.

Our studies can be extended in many different directions: The role of integrability, which the XXZ Hamiltonian possesses, in giving rise to the observed behavior should be explored.   This can be done via the addition of next-nearest neighbor terms that break integrability in numerical simulations, as well as in experiments where such terms can be engineered through appropriate Rydberg dressing of atoms. How would the dynamics change if the initial state is no longer a pure state, but has finite temperature, or if it consists of a single domain wall  \cite{PhysRevE.59.4912, PhysRevE.71.036102, 10.21468/SciPostPhys.7.2.025}? Do the power law scalings change for very large wavelengths which approach the continuum limit?  A preliminary theoretical analysis \cite{Wen_Wei_footnote} suggests this. Experimentally, we can realize Heisenbeg models with purely ferromagnetic couplings by changing the sign of $J_{xy}$ using a constant force to tilt the lattice \cite{tiltedMottInsulator}; we can study the decay of transverse spin via transport and dephasing; we can also explore spin dynamics in two or three spatial dimensions and with higher spin quantum numbers.

\textbf{Acknowledgements} We thank Mikhail Lukin, Norman Yao and Michael Knap for useful discussions, Eunice Lee for experimental assistance, as well as Christoph Paus for sharing computing resources, and Julius de Hond for comments on the manuscript. We acknowledge support from the NSF through the Center for Ultracold Atoms and Grant No. 1506369, ARO-MURI Non-Equilibrium Many-Body Dynamics (Grant No.~W911NF-14-1-0003), ARO-MURI Photonic Quantum Matter (Grant No.~FA9550-16-10323),  AFOSR-MURI Quantum Phases of Matter (Grant No. FA9550-14-10035),  ONR (Grant No. N00014-17-1-2253), the Vannevar-Bush Faculty Fellowship, and the Gordon and Betty Moore Foundation EPiQS Initiative Grant No.~GBMF4306.

\bibliographystyle{naturemag}
\bibliography{spintransport}

\pagebreak

\section{Methods}
\textbf{Extended Hubbard model.}
To determine the parameters $J_{xy}$ and $J_{z}$ in Eq.\,\eqref{Heisenberg_eq}
\begin{align}
    J_{xy} = - \frac{4t^2}{U_{\uparrow\downarrow}}, \qquad J_z = \frac{4t^2}{U_{\uparrow\downarrow}}  - \left( \frac{4t^2}{U_{\uparrow\uparrow}} + \frac{4t^2}{U_{\downarrow\downarrow}}\right) \nonumber
\end{align}
we use measurements of the lattice depth $V_0$ and of the three scattering lengths $a_{\uparrow\uparrow}$, $a_{\uparrow\downarrow}$, $a_{\downarrow\downarrow}$. From the calibrated lattice depth, the Hubbard parameters $t^{(0)}$ (in the non-interacting limit) and $U^{(0)}$ (in the single-band approximation) are calculated as \cite{jacksh98}: 
\begin{equation*}
    t^{(0)} = \int dz\, w^*({z}-{a})\left[ -\frac{\hbar^2}{2 m}\frac{d^2}{dz^2} - V_0\sin^2(kz) \right] w({z})
\end{equation*}
\begin{equation*}
    U^{(0)}_{\sigma \sigma^\prime} = g_{\sigma \sigma^\prime} \int d^3r \, |w({\bf r})|^4
\end{equation*}
where $g_{\sigma \sigma^\prime}\,{=}\,{4\pi\hbar^2 a_{\sigma \sigma^\prime}/m}$ and $\sigma,\sigma^\prime\,{=}\uparrow,\downarrow$, using the calculated lowest-band Wannier functions $w(z)$ for each lattice depth \cite{greinerthesis03}. 

For the precision needed to compare experiment to theory, three corrections are applied \cite{mateusz13, munich_spinHelix,will10,Lhmann2012}. (1) Tunneling is modified by the so-called single-band bond-charge \cite{Lhmann2012}, which for single-occupancy is
\begin{equation*}
    t_{\sigma\sigma^\prime} = t^{(0)} - g_{\sigma \sigma^\prime} \int d^3r \ w^*({\bf r}-\delta{\bf r})w^*({\bf r})w({\bf r})w({\bf r}) \nonumber
\end{equation*}
where $\delta{\bf r}\,{=}\,(0,0,a)$ is a displacement by one lattice constant $a$ in the tunneling direction. Through this correction, the tunnelling matrix elements $t_{\uparrow\uparrow}$, $t_{\uparrow\downarrow}$, $t_{\downarrow\downarrow}$ are now slightly spin-dependent. (2) For the on-site interaction, we include admixtures of higher bands, as discussed in \cite{will10,Lhmann2012}. The dominant part is captured by a perturbative correction due to the first and second excited bands
\begin{align}
    U_{\sigma\sigma^\prime} = U^{(0)}_{\sigma\sigma^\prime} -g_{\sigma\sigma^\prime}^2\sum_{{\bf n}_1,{\bf n}_2}\frac{|\braket{{\bf n}_1,{\bf n}_2|{\bf 0},{\bf 0}}|^2}{E_{\text{bg}}} \nonumber
\end{align}
where ${\bf n}_1$ and ${\bf n}_2$ are the three-dimensional band indices of the two atoms and $E_{\text{bg}}$ is the sum of the bandgap energies. Corrections to the tunneling rate $t$ due to population of higher bands is negligible for a Mott insulator with occupation $n=1$, since higher bands are admixed only through virtual doubly-occupied sites. These modifications of $t$ and $U$ both contribute to a modification of superexchange $J_{xy}\,{=}\,t^2/U$ (Extended Data Fig.~\ref{ED-fig:corrections}). The relative correction to $J^{(0)}_{xy}\,{=}\,(t^{(0)})^2/U^{(0)}$, given by $(J_{xy}-J^{(0)}_{xy})/J^{(0)}_{xy}$, is almost independent of lattice depth (for the range of lattice depths studied), is approximately quadratic in scattering length, and in the experiment $J_{xy}$ is typically reduced by $10\,\%$ to $15\,\%$. Note that the correction has the opposite sign as in \cite{munich_spinHelix} since $J_{xy}$ is anti-ferromagnetic in this work. (3) In order to accurately determine $J_z$, one must also consider off-site interactions of the form \cite{Lhmann2012}
\begin{align}
    V_{\sigma\sigma^\prime}=g_{\sigma\sigma^\prime} \int d^3r \ w^*({\bf r}-\delta{\bf r})w^*({\bf r})w({\bf r}-\delta{\bf r})w({\bf r}), \nonumber
\end{align}
where $\delta{\bf r}$ is defined as before. One finds that $J_z$ is modified by the addition of $ 2 ( V_{\uparrow\uparrow} + V_{\downarrow\downarrow} - 2 V_{\uparrow\downarrow} ) $. Depending on the signs and magnitudes of the three interactions, the off-site terms can add to or subtract from the two previously-discussed corrections to $J$ (Extended Data Fig.~\ref{ED-fig:corrections}).

\textbf{Determination of the Heisenberg parameters.}
We calibrate the lattice depth using amplitude modulation spectroscopy \cite{stofferle04}. We record the excitation spectrum of a Bose-Einstein condensate in a 1D lattice when the depth of the lattice is modulated by $3\,\%$ providing the cloud averaged lattice depth with a statistical uncertainty of of $0.2\,\%$. Due to an asymmetric excitation profile, we estimate a systematic error of $1\,\%$. The trapping potential causes inhomogeneity of the lattice depth (and hence the Heisenberg parameters) across the atom cloud. In our case, the trapping potential is determined solely by the Gaussian curvature of the lattice beams. For a Mott insulator of 44 lattice sites in diameter and lattice beams with $1/e^2$-radius of $125\,\mu\text{m}$, the lattice depth along $z$ varies by $1.3\,\%$. Since in our experiment the lattice depth of the beams along $x$ and $y$ are kept at $V_0\,{=}\,35\,E_R$, this results in a variation at $11\,E_R$ of $t$ by $3.6\,\%$, of $U$ by $0.4\,\%$, and of $J\,{\sim}\,t^2/U$ by $8\,\%$.

The lowest and second-lowest hyperfine states of $^7$Li realize the $\ket{\downarrow}$ and $\ket{\uparrow}$ states. We use our previous measurements \cite{interactionSpectroscopy} of $U_{\uparrow\uparrow}$ and $U_{\downarrow\downarrow}$ (lattice depth modulation) as well as measurements of $U_{\uparrow\uparrow}\,{-}\,U_{\uparrow\downarrow}$ and $U_{\uparrow\downarrow}\,{-}\,U_{\downarrow\downarrow}$ (interaction spectroscopy) to determine the three scattering lengths $a_{\uparrow\uparrow}$, $a_{\uparrow\downarrow}$, $a_{\downarrow\downarrow}$ (under inclusion of higher-band corrections) for several magnetic fields $B$. The determined anisotropies $\Delta$ are shown in Fig.~\ref{fig:setup}b (points). Hyperbolic fits to $a_{\uparrow\uparrow}(B)$, $a_{\uparrow\downarrow}(B)$, $a_{\downarrow\downarrow}(B)$ are used to interpolate the values for the anisotropy (solid line). Extended Data Fig.~\ref{ED-fig:corrections} shows $J_{xy}$, $J_z$ and $\Delta$ with (and without) corrections. A recent detailed theoretical analysis \cite{secker19} of the interaction spectroscopy data also provided precise scattering lengths across several Feshbach resonances.  However, this analysis slightly disagreed with our lattice depth modulation data in the range of magnetic fields studied here, and therefore we relied on the experimental data.

The lattice depth calibration and the experimental determination of the scattering length $a_{\uparrow\downarrow}$ lead to an uncertainty for the spin-exchange times $\hbar/J_{xy}$ of about $\pm10\,\%$. The accuracy of the determined anisotropies $\Delta$ is limited by the experimental determination of all scattering lengths $a_{\uparrow\uparrow}$, $a_{\uparrow\downarrow}$, $a_{\downarrow\downarrow}$. The uncertainty of $\Delta$ is estimated to be about $\pm0.1$.

In the experiment (see Fig.~\ref{fig:setup}b), $\Delta\,{\approx}\,0$ was realized by tuning the magnetic field to $B_0\,{=}\,882.63\,\text{G}$. Here the measured Hubbard parameters result in $\Delta\,{=}\,{-0.12}$ including higher-order correction (and $\Delta\,{=}\,{-0.02}$ without corrections). $B_0^\prime = 842.95\,\text{G}$ is a second magnetic field value, which also realizes $\Delta\,{\approx}\,0$ (actually $\Delta\,{=}\,{-0.13}$ including corrections and $\Delta\,{=}\,0.01$ without corrections). We directly compare these two points $B_0$ and $B_0^\prime$ in Extended Data Fig.~\ref{ED-fig:decay_anisotropies}a~and~b, and observe quantitative agreement. Arbitrary anisotropies were realized by using the magnetic field region in between: $B_0^\prime\,{<}\,B\,{<}\,B_0$. In particular, regions with $\Delta\,{>}\,0$ ($\Delta\,{<}\,0$) were realized by magnetic fields $B\,{<}\,850\,\text{G}$ ($B\,{>}\,850\,\text{G}$). The isotropic point $\Delta\,{\approx}\,1$ was realized at $B_1\,{=}\,847.30\,\text{G}$ (actually $ \Delta\,{=}\,1.01$ and corrections here are negligible). 

\textbf{Experimental Setup.} 
In the experiment, we prepare $4.5\,{\times}\,10^4$ $^7$Li atoms in an optical lattice with spacing $a\,{=}\,532\,\text{nm}$ in the Mott insulating regime with one atom per site \cite{tiltedMottInsulator}. We prepare a far from equilibrium initial spin state and probe the spin dynamics in 1D. The lattice beams in the $x$- and $y$-directions are kept at a large constant value (lattice depth of $35\,E_R$) separating atoms into an array of independent 1D chains, with a typical maximum length of $L_\text{max}\,{=}\,44\,a$ (given by the diameter of the Mott insulator), and with an average length ${\langle L \rangle}\,{=}\,33\,a$ (Extended Data Fig.~\ref{ED-fig:chain_lengths}). Initially, the z-lattice depth is also $35\,E_R$, so that the spin couplings are off. The magnetic field is then ramped to the value required for a desired anisotropy $\Delta$. Using radio frequency pulses and a magnetic field gradient, a helical spin pattern is created where the spin component along the chain winds in the xz-plane of the Bloch sphere with a wavevector $Q\,{=}\,2\pi/\lambda$, where $\lambda$ is the wavelength of the spin helix (see Fig.~\ref{fig:setup}c and Methods section \enquote{Preparation of the spin helix}). For $\lambda$ smaller than the system size, the total magnetization of this state is close to zero.

The power in the third lattice beam ($z$-direction) controls the superexchange rate within the chains. Time evolution is  initiated by ramping down the z-lattice depth to a value between $9$ and $13\,E_R$. The ramp time is $0.5\,\text{ms}$, fast compared to superexchange $\hbar/J_{xy}$, but slow compared to tunneling $\hbar/t$. The ensuing coherent dynamics along each chain is governed by a 1D Heisenberg XXZ-model with anisotropy $\Delta$, Eq.~\eqref{Heisenberg_eq}. This is a quantum quench to a far from equilibrium initial state. After a variable evolution time $t$ the dynamics is frozen by rapidly increasing the lattice depth back to $35\,E_R$. The atoms are imaged in the $\ket{\uparrow}$ state via state-selective polarization-rotation imaging (see below).

\textbf{Preparation of the spin helix.} 
A global $\pi/2$-pulse of $75\,\mu\text{s}$ rotates the spin $\ket{\uparrow}_i$ on each site $i$ into the xy-plane of the Bloch sphere ${\ket{\varphi}_i}\,{=}\,{[\ket{\uparrow}_i+\ket{\downarrow}_i]/\sqrt{2}} $. A magnetic field gradient in z-direction causes spin precession at rates which depend linearly on position $z_i$ of the spin thus creating a spin helix ${\ket{\varphi}_i}\,{=}\,{[\ket{\uparrow}_i+e^{-iQz_i} \ket{\downarrow}_i]/\sqrt{2}} $ where the spin winds in the xy-plane. The strength and duration of the gradient determine the wavevector $Q\,{=}\,2\pi/\lambda$ where $\lambda$ is the wavelength of the spin helix. An additional $\pi/2$-pulse rotates the spin helix into a state where the spin winding occurs in the xz-plane ${\ket{\varphi}_i}\,{=}\,{\sin(Q z_i / 2) \ket{\uparrow}_i}+{\cos(Q z_i / 2) \ket{\downarrow}_i}$, so that the full many-body xz-spin helix state is ${|\psi(Q)\rangle}\,{=}\,{\prod_i \ket{\varphi}_i}$. In practice the phase $\theta$ of the winding varies from realization to realization, which amounts to replacing $Qz_i\,{\mapsto}\,{Q z_i}\,{+}\,\theta$. This is caused by small magnetic bias field drifts on the $10^{-5}$ level. The range of $\lambda$ used in the experiment was limited on the short side by optical resolution to $\lambda\,{\geq}\,5.6\,a$ and on the long side by the length of the chains $L_\text{max}\,{=}\,44\,a$. 

\textbf{Imaging.} 
The optical density of the atomic ensemble is too high ($>14$) to allow for in-situ observation of the modulation of $\braket{S^z}$ via absorption imaging. Instead, we employ dispersive imaging, which uses the phase accumulated by the transmitted light in order to form an image of the atomic density distribution. When light at frequency $\omega_L$ is detuned from the atomic resonance $\omega_0$ by many natural linewidths $\Gamma$, it picks up an approximate phase $\theta\,{\approx}\,{-2\delta/\Gamma\,{\times}\,\text{OD}(x,y,\delta)}$, where OD is the optical density at detuning $\delta\,{=}\,\omega_L\,{-}\,\omega_0$, while absorption is suppressed by a sufficiently large detuning $\delta$. In order to form an image, the phase-shifted light must be interfered with a reference beam. In this work, we make use of the fact that the optical transition we use for imaging is driven only by a single polarization component; after passing through the atoms, the shifted and unshifted components are combined on a polarizer. A judicious choice of input and output polarizers yields an interference signal $I$ on the camera which is $I\,{=}\,I_0(1\,{-}\,\sin\theta)/2$ \cite{huletBEC, ketterle_varenna1999}. 

The optical resolution of our imaging system ($\text{NA}\,{\approx}\,0.2$) was determined to have a cut-off at modulation wavelength $\lambda\,{\approx}\,3.0\,\mu\text{m}\,{=}\,5.6\,a$ (330 line pairs per mm). The reduction of the modulation transfer function $\text{MTF}(Q)$ near the cut-off reduces the observed contrast $\mathcal{C}(t)\,{=}\,\text{MTF}(Q)\,{\cdot}\,c(t)$ compared to the real contrast $c(t)$. This does not affect the decay times $\tau$. Assuming that the experimental preparation sequence for the initial spin helix state achieves full contrast $c(0)\,{=}\,1$ for any wavevector $Q$ (based on careful pulse calibration and characterization), we can use $\mathcal{C}(0)$ as a direct measurement of $\text{MTF}(Q)$ and determine the real contrast as $c(t)\,{=}\,\mathcal{C}(t)/\mathcal{C}(0)$.

\textbf{Constant background contrast.} 
For long evolution times, the contrast does not fully decay, but it asymptotically goes to a finite background value. For example, in Fig.~\ref{fig:XX}a this is about $c_0\,{=}\,0.08$. The fitting function $c(t)\,{=}\,{(a_0\,{+}\,b_0\cos\omega t) e^{-t/\tau}}\,{+}\,c_0$  needs to include this offset $ c_0 $, in order to describe the data accurately. The numerical simulations, however, show a decay to zero for long evolution times $t$. Therefore an offset of $c_0\,{=}\,0.08$ had to be manually added in as well, to see agreement between experiment and theory.
In the experiment, the offset is caused by the inhomogeneous density of the atom cloud:  only $90\,\%$ of the atoms are in the Mott insulator state which realizes an  array of 1D spin chains. A small fraction of atoms are in dilute spatial wings, separated by holes which are immobile due the gradient of the trapping potential (which suppresses first-order tunneling as shown in our previous work \cite{tiltedMottInsulator}). These atoms preserve an imprinted spin modulation pattern for long times. We have checked this mechanism by increasing the amount of thermal atoms and clearly observe an increase of the background  $c_0$ (Extended Data Fig.~\ref{ED-fig:holes}). Furthermore, a position sensitive measurement of the contrast confirms, that the main contribution is indeed from atoms in the spatial wings (Extended Data Figs.~\ref{ED-fig:holes}f and \ref{ED-fig:holes}g). In agreement with this model, numerical simulations always show a decay to zero for long spin chains (see Extended Data Figs.~\ref{ED-fig:universal}, \ref{ED-fig:negative_anisotropies}, \ref{ED-fig:initial_phase}, \ref{ED-fig:chain_lengths}).

\textbf{Time-reversal invariance of spin dynamics.}
For an xz-spin helix initial state and time evolution via the XXZ Hamiltonian, the contrast is time-reversal symmetric: $c(t)\,{=}\,c(-t)$, which follows because the state, Hamiltonian and observable (the local magnetization $S^z_i$) can be all expressed real. This also implies invariance against the overall sign of the Hamiltonian $H\,{\mapsto}\,{-H}$. The same argument holds for the system with holes evolving under the bosonic $t$-$J$ model. The initial dynamics of the contrast (in the ideal scenario) in both cases is therefore quadratic, $c(t)\,{=}\,1\,{-}\frac{1}{2}\Gamma^2 t^2\,{+}\,\cdots$, with 
\begin{equation*}
    \Gamma^2 = -\frac{2}{L} \sum_i  \cos(Qz_i + \theta) \langle \psi(Q) | [H,[H,S^z_i] | \psi(Q)\rangle
\end{equation*}
Therefore the timescale of the initial quadratic decay $|\Gamma|^{-1}$ is the superexchange timescale $\hbar/J_{xy}$ (XXZ model) or $\hbar/t$ ($t$-$J$ model). The fact that experimentally we do not observe an initial quadratic behavior indicates either (i) the presence of holes, but we are not resolving the fast timescale $\hbar/t$, or (ii) that the initial state is not time-reversal invariant (i.e.~cannot be expressed real, in the same basis that  the Hamiltonian is written in). The latter could arise from pulse imperfections, or the fact that the ramp-down of the optical lattice takes place over a finite duration of time, leading to deviations from the ideal initial state. Nevertheless, we expect that the overall behavior of the  decay of the visibility, e.g.~scaling behavior of dynamics with wavevector $Q$, is not strongly affected by (i) or (ii).
 
\textbf{Numerical simulations.} 
In the numerical simulations we consider: (i) a spin helix quench under XXZ Hamiltonian (Eq.~\eqref{Heisenberg_eq}) dynamics, and (ii) a spin helix with $2.5\,\%$ to $5\,\%$ hole probability evolving under the bosonic $t$-$J$ model (i.e. assuming no doubly occupied sites), given by
\begin{align*}
    H & = \sum_{\langle ij \rangle } J_{xy}(S^x_i S^x_j + S^y_i S^y_j) + J_z S^z_i S^z_j  + H_d \nonumber \\
\end{align*}
with
\begin{align*}
    H_d & = -\sum_{\sigma,\langle ij \rangle } t a_{\sigma i}^\dagger a_{\sigma j} - \sum_{\sigma, \langle i jk \rangle} \Bigg[\frac{t^2}{U_{\uparrow \downarrow}} a_{\sigma i}^\dagger n_{\bar\sigma j} a_{\sigma k}    \nonumber \\ 
    &  + \frac{t^2}{U_{\uparrow \downarrow}} a_{\bar\sigma i}^\dagger S_j^\sigma a_{\sigma k} + \frac{2t^2}{U_{\sigma\sigma}} a_{\sigma i}^\dagger n_{\sigma j} a_{\sigma k}+ \text{h.c.} \Bigg]
    \label{eqn:tJ}
\end{align*}
where spin ${\sigma}\,{=}\,{\uparrow,\downarrow}$. Here $a_{\sigma i}$, $a^\dagger_{\sigma i}$ are bosonic lowering and raising operators at site $i$,  $S_i^\uparrow\,{\equiv}\,S_i^+$ is defined as $a_{\uparrow i}^\dagger a_{\downarrow i}$, while $S_i^\downarrow\,{\equiv}\,S_i^-$ is defined as $a_{\downarrow i}^\dagger a_{\uparrow i}$. 
We use parameters from experiments and focus on a lattice depth of $11\,E_R$, in which case we have $U_{\uparrow\downarrow}/U_{\uparrow\uparrow}\,{=}\,1.206$, $1.406$, $1.401$, $1.398$, $1.397$, $1.392$, $U_{\uparrow\downarrow}/U_{\downarrow\downarrow}\,{=}\,{-0.188}$, $ 0.264$, $0.459$, $0.575$, $0.659$, $0.862$ and $U_{\uparrow\downarrow}/t\,{=}\,{-17.94}$, $-24.32$, $ -24.17$, $-24.08$, $-24.05$, $-23.94$ for anisotropies $\Delta\,{=}\,0.020$, $0.670$, $0.860$, $0.973$, $1.055$, $1.256$ respectively. In the absence of holes, the action of the term $H_d$ vanishes, and the Hamiltonian reduces to the XXZ Hamiltonian.

In both cases we employ the real-time time-evolving block decimation (TEBD) method with matrix product states (MPS) on an open spin chain. In case (i): local Hilbert-space dimension$\,{=}\,2$, length $L\,{=}\,40$ sites and bond dimensions up to $800$; case (ii):  local Hilbert-space dimension$\,{=}\,3$, $L\,{=}\,40$ for $\Delta\,{=}\,0$ or $L\,{=}\,20$ for all other $\Delta$. Simulations are cut off in simulated time due to the rapid increase in entanglement of the state, requiring ever increasing computational times. For the special case $\Delta\,{=}\,0$, without holes, we alternatively employ free fermionic methods, significantly speeding up the calculations. For technical reasons, instead of simulating the full distribution of holes we simply average over the situations in which there are either exactly one or two holes in the chain. For $L\,{=}\,20$ ($40$), a single hole corresponds to an average of $5\,\%$ ($2.5\,\%$) holes.

To obtain the contrast, we determine the local magnetization $\langle S^z_i(t) \rangle\,{=}\,\langle n_{i,\uparrow}(t) \rangle\,{-}\,\langle n_{i,\downarrow}(t) \rangle$ and determine the Fourier component at wavevector $Q$ via $c(t)\,{:=}\,\frac{2}{L} \sum_i \langle S^z_i(t) \rangle \cos(Q z_i + \theta)$, assuming that the initial spin helix has wavevector $Q$ with a given  phase $\theta$ (see Methods section \enquote{Creation of the spin helix}); we then compare this to the experimentally extracted contrast. However as noted in the Methods section \enquote{Time-reversal invariance of spin dynamics}, the numerically simulated contrast always has an initial quadratic decay, unlike that seen in the experiments. We therefore determine the decay timescales and its power law scaling with $Q$ with one of the following methods: for positive anisotropies $\Delta$, we define the decay time constant as the time it takes for the contrast $c(t)$ to decay from 1 to 0.6, multiplied by ${-\ln(0.6)}$ to convert to a \enquote{$1/e$} time (see Extended Data Fig.~\ref{ED-fig:universal}). For negative anisotropies, for regime I, we take the time to decay from 1 to 0.9, while for regime II we fitted a simple exponential decay profile to obtain the decay timescales  once the curves start \enquote{peeling off} when plotted in time rescaled by $\lambda$ (see Fig.~\ref{fig:negative}c and Extended Data Fig.~\ref{ED-fig:negative_anisotropies}). The resulting power law scalings (Extended Data Fig.~\ref{ED-fig:powerlaw_theory}) and resulting exponents (Fig.~\ref{fig:powerlaw}c) are in reasonable agreement with the experimental results.

\textbf{Finite size effects.}
Numerical simulations can study the effect of different chain lengths and initial phases of the spin helix on the dynamics. This is relevant because in the experiment, the atoms are distributed uniformly over a sphere with a typical diameter of $L_\text{max}\,{=}\,44\,a$ in three-dimensions, leading to an ensemble of 1D chains of varying lengths. The experimentally measured contrast $c(t)$ is an average over all chain lengths with a probability distribution shown in Extended Data Fig.~\ref{ED-fig:chain_lengths}. Furthermore, due to drifts of the applied magnetic field, the initial phase $\theta$ of the spin helix state varies from shot to shot. Here we study numerically both effects.

We concentrate on the XX model ($\Delta\,{=}\,0$) without holes.  Extended Data Fig.~\ref{ED-fig:initial_phase} shows that the strong dependence of the  visibility $c(t)$ on the initial phase is due to reflection of magnetization off the boundaries of the chain. This suggests that averaging over various chain lengths or averaging over initial phases should give similar results which is confirmed in Extended Data Fig.~\ref{ED-fig:chain_lengths}.  It is even sufficient to average over only two phases, $\theta\,{=}\,0$ and $\pi/2$, to achieve insensitivity to initial and boundary conditions. The distribution of chains, and the simulation of a magnetization profile averaged over the distribution of chains, are illustrated in Extended Data Fig.~\ref{ED-fig:chain_lengths}b-d.

The conclusion is that the experiment is naturally performing an average over different phases and different chain lengths, washing out the sensitive dependence of the spin dynamics on initial conditions.  For comparison with simulations, it is sufficient to use a system with a  fixed chain length $L\,{=}\,40\,a$, and average over only the two phases $\theta\,{=}\,0$ and $\pi/2$.

\setcounter{figure}{0}
\renewcommand{\figurename}{Extended Data Fig.} 

\onecolumngrid

\newpage

\begin{figure}[h]
    \includegraphics[width=0.8\linewidth,keepaspectratio]{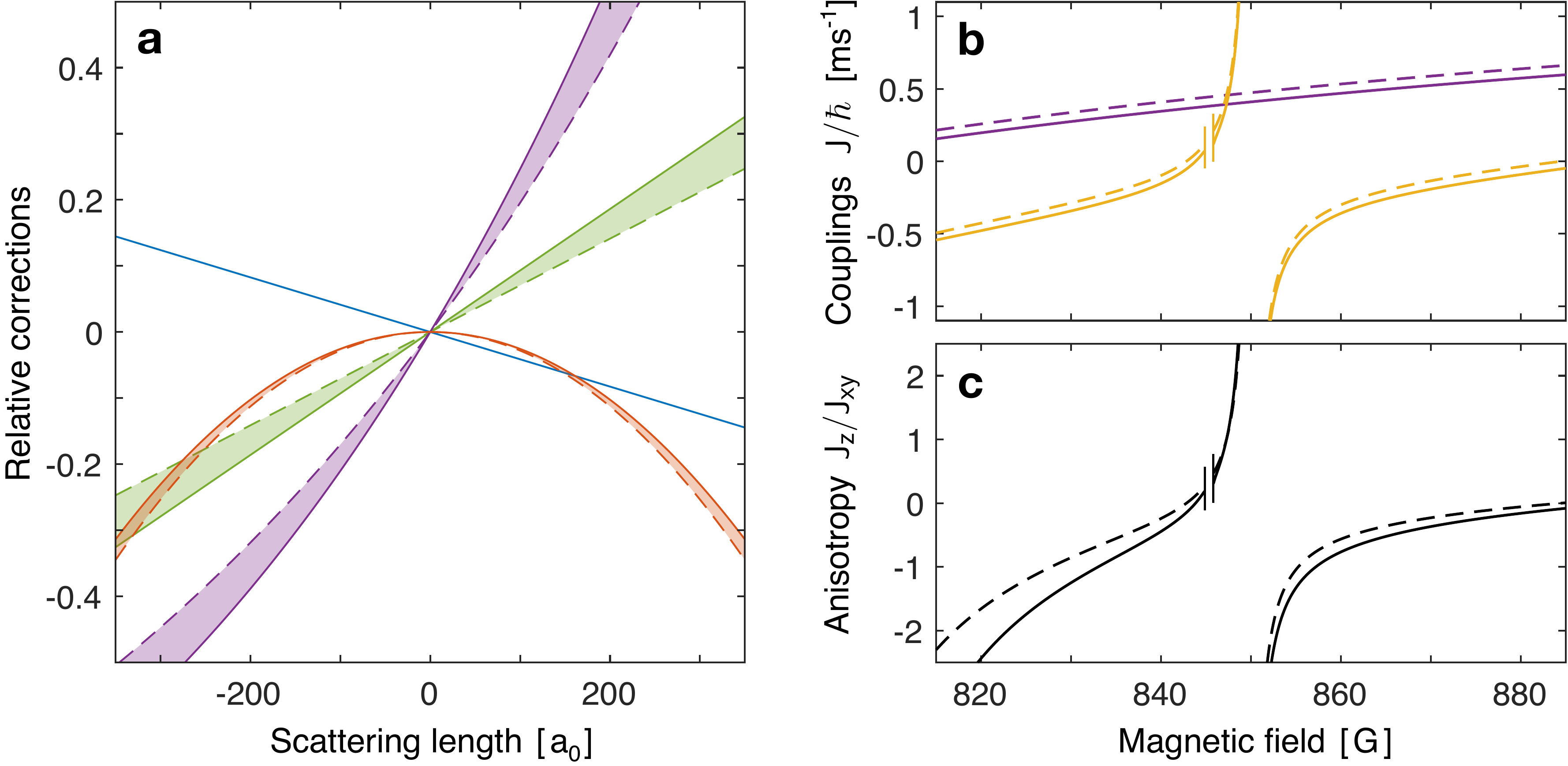}
    \caption{
        \textbf{Corrections to Hubbard parameters}. \textbf{a}, Corrections for tunnelling $ (t-t^{(0)})/t^{(0)} $ (green), on-site interactions $ (U-U^{(0)})/U^{(0)} $ (blue), superexchange $ (J-J^{(0)})/J^{(0)} $ (purple) and off-site interactions $ -2 V / J^{(0)} $ (red), where $ t^{(0)} $, $ U^{(0)} $, $ J^{(0)} = 4(t^{(0)})^2/U^{(0)}$ are the uncorrected values and $ t $, $ U $, $ J = 4t^2/U $ include corrections, at a lattice depth of $13\,E_R$ (solid line) and $9\,E_R$ (dashed line). \textbf{b},\textbf{c}, As a function of magnetic field $ B $ we show the transverse coupling constant $J_{xy}$ (yellow), the longitudinal coupling constant $J_{z}$ (purple) and the anisotropy $\Delta\,{=}\,J_z/J_{xy} $ (black), without corrections (dashed) and including corrections (solid) for a lattice depth of $11\,E_R$. The excluded region ($|a_{\uparrow\uparrow}|\,{>}\,700\,a_0$) is around a Feshbach resonance in the $\ket{\uparrow}$ state near $845.4\,\text{G}$.}
	\label{ED-fig:corrections}
\end{figure}

\newpage

\begin{figure}[h]
    \includegraphics[width=1\linewidth,keepaspectratio]{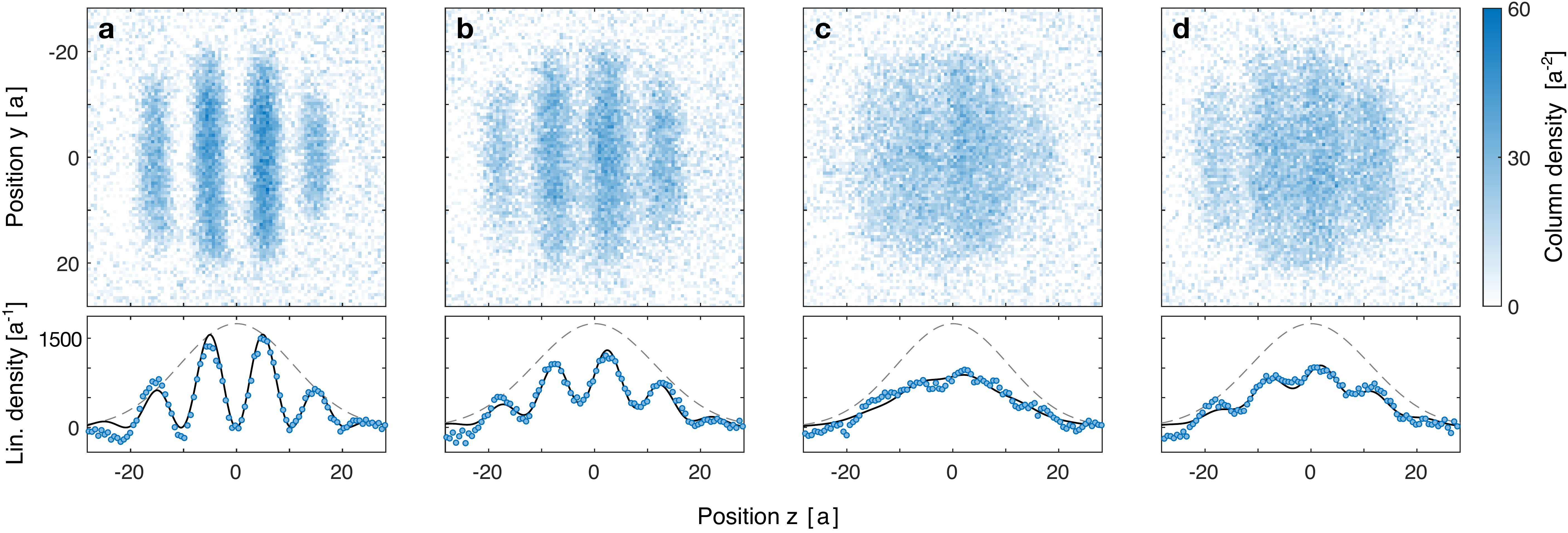}
    \caption{
        \textbf{Contrast measurement}. The images \textbf{a}-\textbf{d} show the distribution of atoms in the $\ket{\uparrow}$ state. Every pixel is a local measurement of the column density (number of atoms per unit area). The y- and z-axes are displayed in units of lattice spacings $a\,{=}\,0.532\,\mu\text{m}$. The images are projected (integrated) along the y-direction to obtain the linear density (number of atoms per unit length). The resulting 1D distributions are fitted with $f(z)\,{=}\,{g(z)\,{\cdot}\,[{1}\,{+}\,{\mathcal{C} \cos(Qz+\theta)}]/2}$ (solid line) where $ g(z) $ is a Gaussian envelope (dashed line). Examples \textbf{a}-\textbf{d} were measured at different evolution times (\textbf{a}) $t\,{=}\,0\,\hbar/J_{xy}$, (\textbf{b}) $t\,{=}\,2.3\,\hbar/J_{xy}$, (\textbf{c}) $t\,{=}\,6.3\,\hbar/J_{xy}$, (\textbf{d}) $t\,{=}\,12.0\,\hbar/J_{xy}$, for anisotropy $\Delta\,{\approx}\,0$ and wavelength $\lambda\,{=}\,10.4\,a$. The obtained contrast $\mathcal{C}(t)$ is shown in Fig.~2a. In general, we also normalize by the initial measured contrast $\mathcal{C}(0)$ to correct for finite optical imaging resolution. This is important for shorter wavelengths $\lambda$ close to the optical resolution of $3\,\mu\text{m}$, where the measured contrast $\mathcal{C}(t)$ is reduced compared to the real contrast $c(t)\,{=}\,\mathcal{C}(t)/\mathcal{C}(0)$. }
	\label{ED-fig:imageFit}
\end{figure}

\newpage

\begin{figure}[h]
    \includegraphics[width=1\linewidth,keepaspectratio]{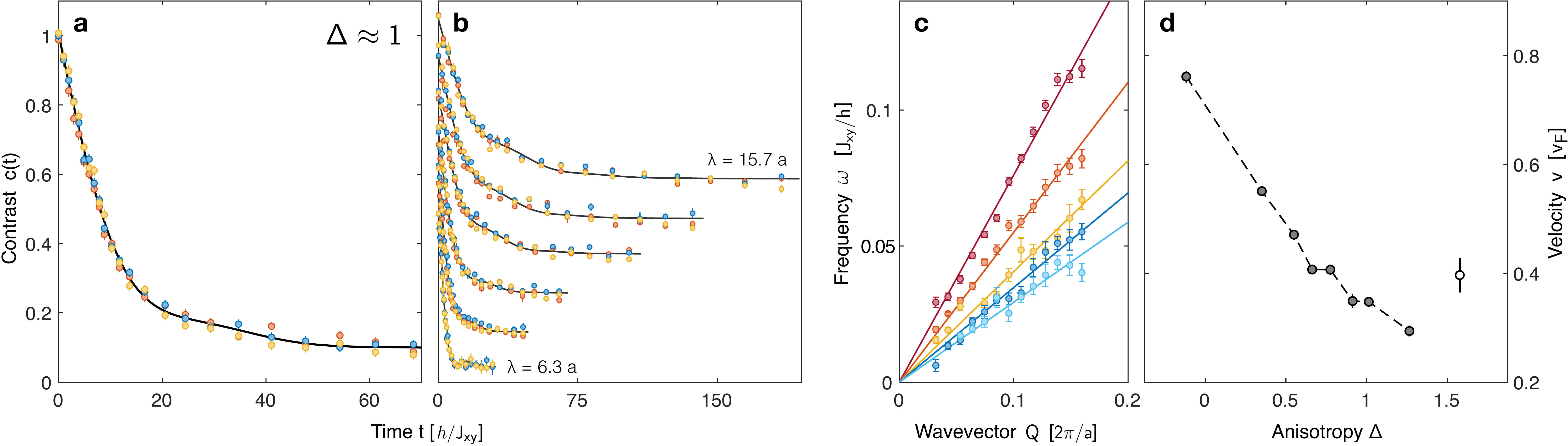}
    \caption{
        \textbf{Dispersion relations}. For all positive anisotropies $\Delta\,{\geq}\,0$, the time evolution of the contrast $c(t)$ shows a damped oscillatory component, in addition to the overall exponential decay. For larger $ \Delta $, the oscillations become smaller. \textbf{a}, Decay and weak oscillation of the contrast  at the isotropic point $\Delta\,{\approx}\,1$. The observed period is $T\,{=}\,2\pi/\omega\,{=}\,37(2)\,\hbar/J_{xy}$ for $ \lambda = 10.4a $. \textbf{b}, Decay curves for different wavelengths $\lambda\,{=}\,15.7\,a$, $13.4\,a$, $11.7\,a$, $9.4\,a$, $7.8\,a$, $6.3\,a$ show how the measured oscillation frequencies $\omega$ vary with $\lambda$. \textbf{c}, They follow linear dispersion relations $\omega(Q)\,{=}\,v Q$ shown for $\Delta\,{=}\,{-0.12}$ (red), $0.35$ (orange), $0.78$ (yellow), $1.01$ (blue) and $1.27$ (light blue). \textbf{d}, The obtained velocities $v$ decrease with increasing anisotropy $\Delta$. For $\Delta\,{=}\,1.58$ (open symbol) oscillations were very small and the measurement was limited to large values of $Q$, which precluded recording a full dispersion relation.}
	\label{ED-fig:dispersion}
\end{figure}

\newpage

\begin{figure}[h]
    \includegraphics[width=1.0\linewidth,keepaspectratio]{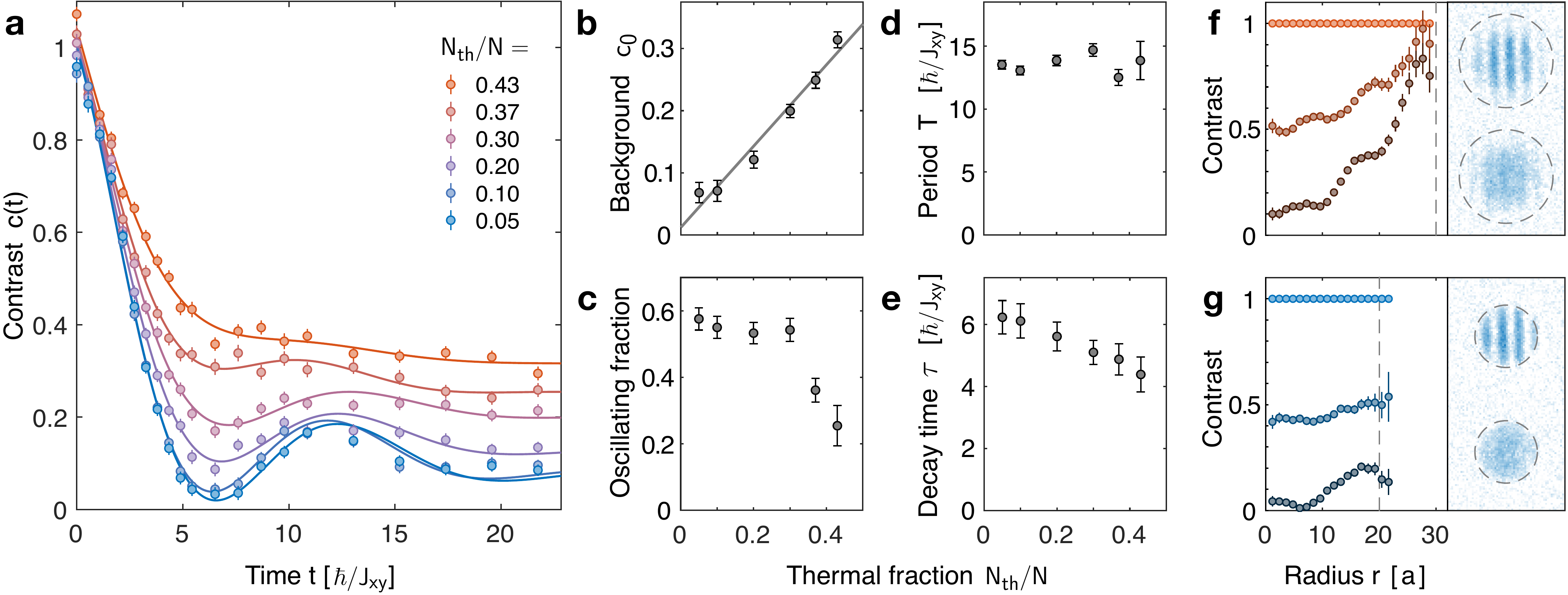}
    \caption{
        \textbf{Effect of finite hole concentration} shown for $\Delta\,{\approx}\,0$ and $\lambda\,{=}\,10.4\,a$.  By varying the thermal fraction $N_\text{th}/N$ of the Bose-Einstein condensate before it is loaded into the optical lattice, we vary the energy and entropy of the atoms in the spin chain, and therefore the concentration of holes. (For our conditions, doubly occupied sites have higher energies than holes).  \textbf{a}, Decay curves $c(t)$ for varying hole concentrations ranging from low (blue) to high (red) thermal fraction. Solid lines are fits $c(t)\,{=}\,\left(a_0\,{+}\,b_0\cos\omega t\right) e^{-t/\tau}\,{+}\,c_0$. \textbf{b}, The background contrast $c_0$ increases monotonously with thermal fraction $N_\text{th}/N$. A linear fit (solid line) extrapolates to $c_0\,{=}\,0.01(2)$, consistent with zero, for $N_\text{th}/N\,{=}\,0$. This suggests that all of the background contrast is due to hole excitations.  Higher hole concentrations suppress (\textbf{c}) the oscillating fraction $b_0/(a_0\,{+}\,b_0)$, (\textbf{d}) don't affect the oscillation period $T\,{=}\,2\pi/\omega$ and (\textbf{e}) decrease the decay time $\tau$ only slightly.  The behavior shown in \textbf{c} and \textbf{e} is most likely caused by mobile holes in the central part of the Mott insulator -- indeed, numerical simulations of the $t$-$J$-model reproduce such effects (Fig.~2a). On the other hand, a finite background contrast (\textbf{b}) is likely caused by immobile holes located in the outer parts of the atom distribution where first-order tunneling is suppressed by the gradient of the (harmonic) trapping potential \cite{tiltedMottInsulator}. Since they disrupt spin transport, we expect that the imprinted spin modulation in these regions will not (or only very slowly) decay. This region is visible as a shell of low atomic density surrounding the Mott insulator in the in-situ images for (\textbf{f}) large hole concentration and absent for (\textbf{g}) low hole concentration. For different evolution times $t$, the plots in \textbf{f} and \textbf{g} show (on the left) the local contrast as function of distance $r$ from the center of the atom cloud, obtained from in-situ images (on the right). From top to bottom, the evolution times are $t\,{=}\,0\,\hbar/J_{xy}$ (top image), $2.7\,\hbar/J_{xy}$ and $21.7\,\hbar/J_{xy}$ (bottom image). The dashed line indicates a contour of constant radius (\textbf{f}) $r\,{=}\,30\,a$ and (\textbf{g}) $r\,{=}\,20\,a$. }
	\label{ED-fig:holes}
\end{figure}

\newpage

\begin{figure}[h]
    \includegraphics[width=0.7\linewidth,keepaspectratio]{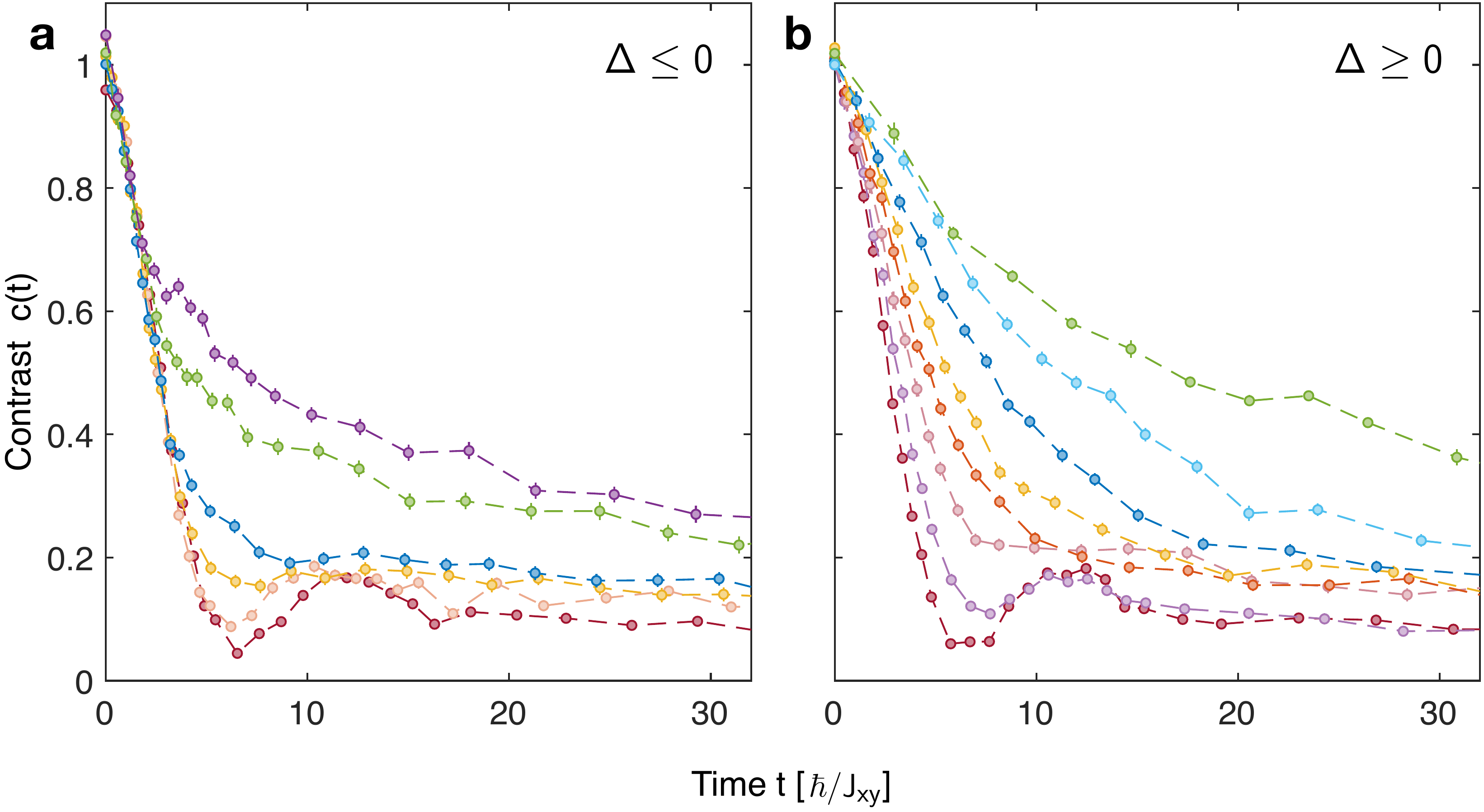}
    \caption{
        \textbf{Decay behavior as a function of anisotropy} ranging from (\textbf{a}) negative to (\textbf{b}) positive, for one fixed wavelength $\lambda\,{=}\,10.4\,a$. Using $\Delta\,{\approx}\,0$ as a reference point, we show how the temporal profile of the decay curve $c(t)$ changes when we introduce positive or negative interactions. Every data point is an average of two measurement at lattice depths $9\,E_R$ and $11\,E_R$. \textbf{a}, from bottom to top: $\Delta\,{=}\,{-0.12}$ (red), ${-0.59}$ (pink), ${-0.81}$ (yellow), ${-1.02}$ (blue), ${-1.43}$ (green), ${-1.79}$ (purple). \textbf{b}, from bottom to top: $\Delta\,{=}\,{-0.13}$ (red), $0.08$ (purple), $0.35$ (pink), $0.55$ (orange), $0.78$ (yellow), $1.01$ (blue), $1.27$ (light blue), $1.58$ (green). Regardless of the sign, for increasing $|\Delta|$ the decay always slows down and the revivals damp out more quickly. However, there is a big difference in how this slowdown happens: \textbf{b}, For increasing positive interactions $\Delta\,{>}\,0$, the initial rate of decay increases continuously. \textbf{a}, In contrast, for all negative interactions $\Delta\,{<}\,0$, the initial rate of decay stays constant (and is ballistic), coinciding with the $\Delta\,{\approx}\,0$ case. It is only after a critical time $t_0$ that the decay suddenly starts slowing down (and becomes diffusive) for times $t\,{>}\,t_0$. This critical time $t_0$ decreases with increasing negative interaction strength $|\Delta|$.}
		\label{ED-fig:decay_anisotropies}
\end{figure}

\newpage

\begin{figure}[h]
    \includegraphics[width=1\linewidth,keepaspectratio]{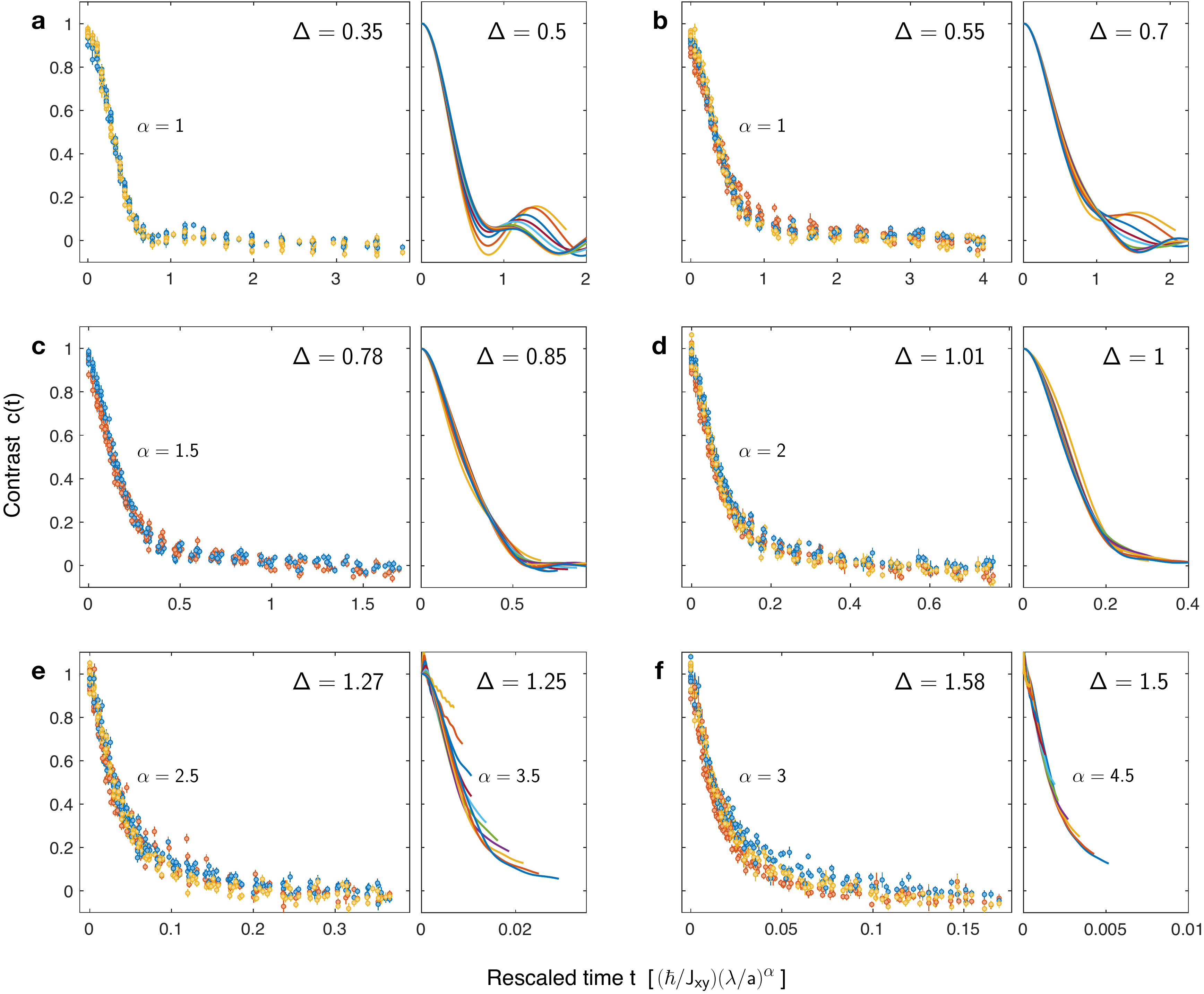}
    \caption{
        \textbf{Collapse of decay curves} for positive anisotropies $\Delta\,{>}\,0$. All decay curves $c(t)$ for different wavelengths $\lambda\,{=}\,15.7\,a$, $13.4\,a$, $11.7\,a$, $10.4\,a$, $9.4\,a$, $8.5\,a$, $7.8\,a$, $7.2\,a$, $6.7\,a$  collapse very well into a single curve for all evolution times $t$, when time units are rescaled by $\lambda^\alpha$, where the exponent $\alpha$ is a function of anisotropy $\Delta$, both for experiment (points) and theory (solid lines). Experimental points were measured for lattice depths: $9\,E_R$ (red), $11\,E_R$ (blue), $13\,E_R$ (yellow). {\textbf{a}, \textbf{b},} Ballistic regime ($\alpha\,{=}\,1$), \textbf{c}, super-diffusion ($\alpha\,{=}\,1.5$), \textbf{d}, diffusion ($\alpha\,{=}\,2$), \textbf{e}, \textbf{f}, sub-diffusion ($\alpha\,{=}\,2.5,\,3$ for experiment and $\alpha\,{=}\,3.5,\,4.5$ for numerical simulations. In \textbf{f}, experiments covered a reduced range $\lambda\,{\leq}\,10.4\,a$). However, the experimentally measured oscillation frequencies $\omega$ still follow a linear dispersion relation for all anisotropies $\Delta\,{\geq}\,0$ (Extended Data Fig.~\ref{ED-fig:dispersion}). Therefore the collapse is not perfect. Nevertheless, away from the ballistic regime $\alpha\,{\approx}\,1$, such oscillations are small and therefore only lead to a small deviation from the collapse behavior. Note also the rather different timescales in experiments and simulations for $\Delta\,{>}\,1$.}
	\label{ED-fig:universal}
\end{figure}

\newpage

\begin{figure}[h]
    \includegraphics[width=1\linewidth,keepaspectratio]{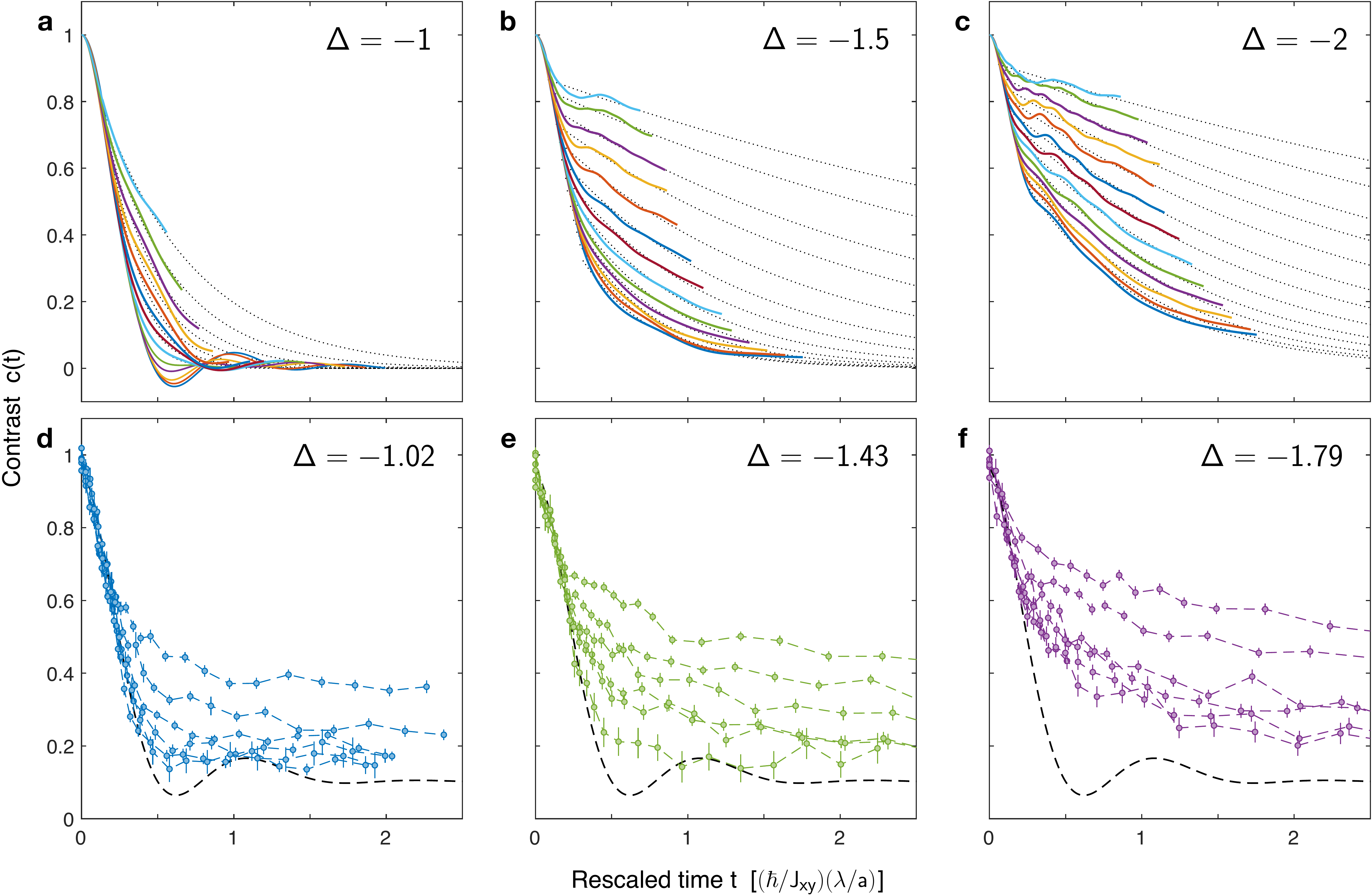}
    \caption{
        \textbf{Collapse at short times} for negative anisotropies $\Delta\,{<}\,0$. All decay curves $c(t)$ for different wavelengths $\lambda$ collapse into a single curve at early times, when time units are rescaled by $\lambda$ (indicating ballistic behavior). For later times the decay is diffusive with different scaling. \textbf{a}-\textbf{c}, Theory (from top to bottom: $\lambda\,{=}\,31.3\,a$, $23.5\,a$, $18.8\,a$, $15.7\,a$, $13.4\,a$, $11.7\,a$, $10.4\,a$, $9.4\,a$, $8.5\,a$, $7.8\,a$, $7.2\,a$, $6.7\,a$, $6.3\,a$). The dotted lines are exponential fits $e^{-t/\tau_\text{II}}$ to the diffusive regime and the time constants $\tau_\text{II}$ are shown in Extended Data Fig.~\ref{ED-fig:powerlaw_theory}a. \textbf{d}-\textbf{f}, Experiment (from top to bottom: $\lambda\,{=}\,18.8\,a$, $13.4\,a$, $10.4\,a$, $8.5a\,$, $7.2\,a$, $6.3\,a$) shown for a lattice depth of $11\,E_R$. The dashed line indicates the ballistic case $\Delta\,{\approx}\,0$ (see Fig.~2c).}
	\label{ED-fig:negative_anisotropies}
\end{figure}

\newpage

\begin{figure}[h]
    \includegraphics[width=0.9\linewidth,keepaspectratio]{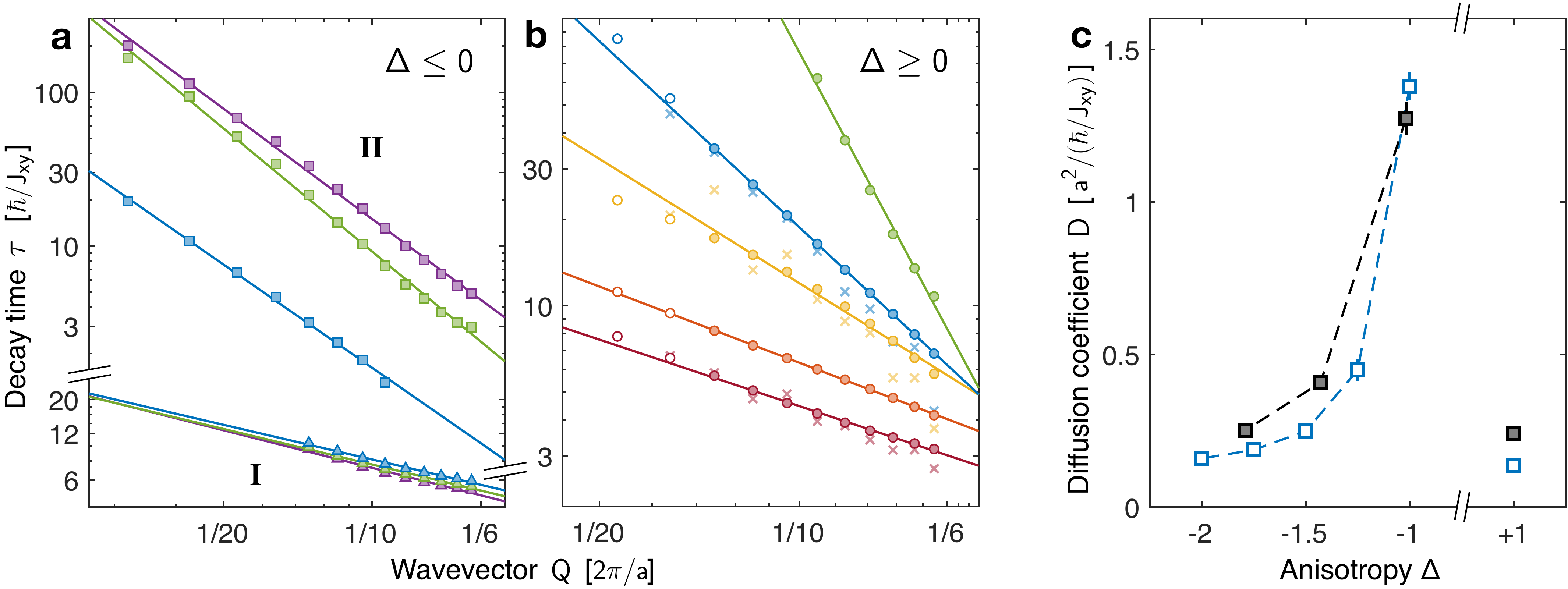}
    \caption{
        \textbf{Power law scalings (theory) and diffusion coefficients}. \textbf{a},\textbf{b}, Decay time constants $\tau$ for different anisotropies $\Delta$ ranging from (\textbf{a}) negative to (\textbf{b}) positive. Numerical results are shown in \textbf{a} for $\Delta\,{=}\,{-1}$ (blue), ${-1.5}$ (green), ${-2}$ (purple) and in \textbf{b} for $\Delta\,=\,0$ (red), ${0.5}$ (orange), ${0.85}$ (yellow), ${1}$ (blue), ${1.5}$ (green). Solid lines are power law fits (to the filled symbols). Open symbols are excluded from the fit due to finite size effects. Crossed symbols are results from $t$-$J$-model simulations including $5\,\%$ hole fraction. Fitted power law exponents are shown in Fig.~4c of the main text. For positive anisotropies $\Delta\,{\geq}\,0$ the decay time $\tau$ is defined as $\tau\,{=}\,\tau'/\ln(1/0.60)$ with $c(\tau')\,{=}\,0.60$. For negative anisotropies $\Delta\,{\leq}\,0$, the decay time $\tau_\text{I}$ for (I) short times is defined as $\tau_\text{I}\,{=}\,10\tau_\text{I}'$ with $c(\tau_\text{I}')\,{=}\,0.90$. For (II) longer times, the decay time $\tau_\text{II}$ is obtained from exponential fits $e^{-t/\tau_\text{II}}$ (see dotted curves in Extended Data Fig.~\ref{ED-fig:negative_anisotropies}). \textbf{c}, Diffusion coefficients obtained from theory (open symbols) and experiment (filled symbols). For negative anisotropies $\Delta\,{<}\,0 $, values were determined from quadratic power law fits $1/\tau\,{=}\,DQ^2$ to the data points in \textbf{a} (theory) and Fig.~3a (experiment) for the diffusive regime (II). Note that for $\Delta\,{\geq}\,0$ the system is only diffusive for $\Delta\,{=}\,{+1}$, as shown in \textbf{b} (theory) and Fig.~3b (experiment). }
	\label{ED-fig:powerlaw_theory}
\end{figure}

\newpage

\begin{figure}[h]
    \includegraphics[width=\linewidth,keepaspectratio]{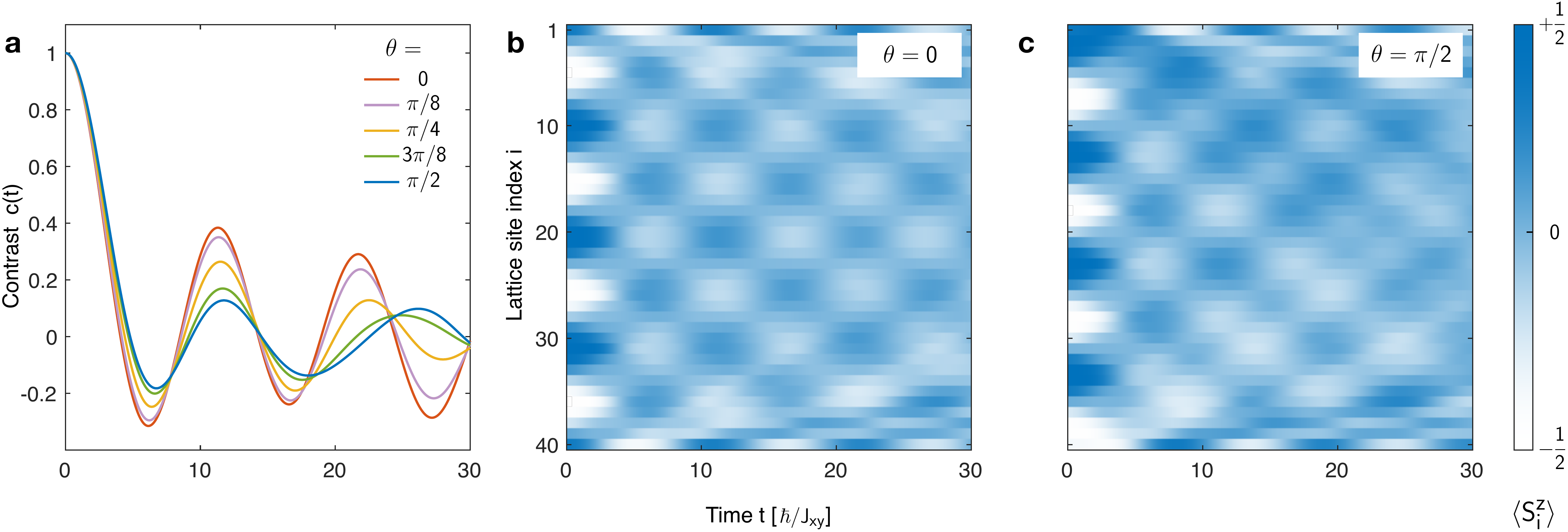}
    \caption{
        \textbf{Finite size effects from the initial phase} of the spin helix state, illustrated here for $\Delta\,{=}\,0$ and $\lambda\,{=}\,10.4\,a$. \textbf{a}, The time evolution of the contrast $c(t)$ depends strongly on the initial phase $\theta$. The dynamics of the local magnetization $\langle S^z_i(t) \rangle$ for phases (\textbf{b}) $\theta\,{=}\,0$ and (\textbf{c}) $\theta\,{=}\,\pi/2$ reveals that this arises due to the reflection of ballistically propagating magnetization off the ends of the chain. Depending on the initial phase of the spin helix, the reflected magnetization interferes constructively or destructively with the  pattern of the bulk magnetization. }
	\label{ED-fig:initial_phase}
\end{figure}

\newpage

\begin{figure}[h]
    \includegraphics[width=\linewidth,keepaspectratio]{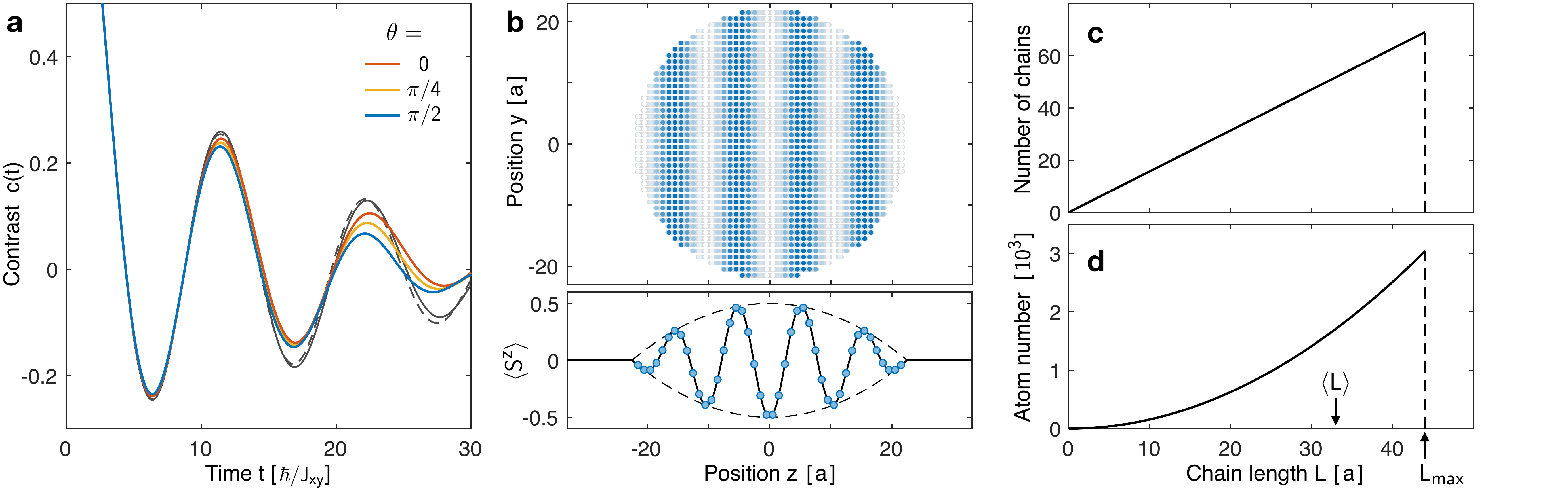}
    \caption{
        \textbf{Finite size effects from the chain length}. \textbf{a}, Contrast $c(t)$ obtained after a weighted average over all different chain lengths between $L = 0$ and $44$, as illustrated in (\textbf{b}), for $\Delta\,{=}\,0$ and $\lambda\,{=}\,10.4\,a$. The dynamics shows almost no dependence on the phase $\theta$ (red, yellow, blue) in contrast to a single chain length $L\,{=}\,40$ (Extended Data Fig.~\ref{ED-fig:initial_phase}a). Also overlaid are the contrast for a fixed chain length ($L\,{=}\,40$) averaged over all initial phases $0\,{\leq}\,\theta\,{<}\,2\pi$ (black solid line), and averaged over only the two phases $\theta\,{=}\,0$ and $\pi/2$ (black dashed line). The close agreement implies that averaging over either chain lengths or phases suppresses the dependence on initial or boundary conditions. \textbf{b}, A cut through the spherical Mott insulator with diameter $L_\text{max}\,{=}\,44\,a$ (as in the experiment) illustrates the distribution of different chain lengths (oriented along the z-direction). Averaging the local magnetization $S^z$ over the x- and y-direction provides a 1D magnetization profile (bottom panel), which is an average over all chains. \textbf{c}, The number of chains with length $L$ is given by $(\pi/2)(L/a)$.  The total number of chains is $\pi(L_\text{max}/2a)^2\,{\approx}\,1500$. \textbf{d}, The number of atoms in chains with length $L$ is given by $(\pi/2)(L/a)^2$. Since the contribution of each chain to the imaging signal is proportional to the atom number in the chain, the relevant average over chain lengths is weighted by the atom number and is $\langle L \rangle\,{=}\,3/4\,{\cdot}\, L_\text{max}\,{=}\,33\,a$. }
	\label{ED-fig:chain_lengths} 
\end{figure}

\end{document}